\documentclass[twocolumn]{aastex631}

\newcommand\kepler{\textit{Kepler}}
\newcommand\gaia{\textit{Gaia}}

\newcommand\numax{$\nu_{\rm max}$}
\newcommand\dnu{$\Delta\nu$}
\newcommand\teff{$T_{\rm eff}$}
\newcommand\metal{[Fe/H]}
\newcommand\amlt{{\ensuremath{\alpha_\mathrm{MLT}}}}

\usepackage{xcolor}
\usepackage{subfigure}
\usepackage[capitalize]{cleveref}
\usepackage{chngcntr}
\usepackage{newunicodechar}

\DeclareRobustCommand{\okina}{%
  \raisebox{\dimexpr\fontcharht\font`A-\height}{%
    \scalebox{0.8}{`}%
  }%
}
\newunicodechar{ʻ}{\okina}

\shorttitle{Detailed Asteroseismic Modeling of KOI-3886 and $\iota$ Draconis}
\shortauthors{Campante et al.}

\graphicspath{{./}{figures/}}

\begin{document}

\title{Revisiting the Red-giant Branch Hosts KOI-3886 and $\iota$ Draconis.\\ Detailed Asteroseismic Modeling and Consolidated Stellar Parameters}

\correspondingauthor{Tiago L. Campante, Tanda Li,\\ J. M. Joel Ong}
\email{tiago.campante@astro.up.pt}
\email{litanda@bnu.edu.cn}
\email{joelong@hawaii.edu}

\author[0000-0002-4588-5389]{Tiago L. Campante}
\affiliation{Instituto de Astrof\'{\i}sica e Ci\^{e}ncias do Espa\c{c}o, Universidade do Porto,  Rua das Estrelas, 4150-762 Porto, Portugal}
\affiliation{Departamento de F\'{\i}sica e Astronomia, Faculdade de Ci\^{e}ncias da Universidade do Porto, Rua do Campo Alegre, s/n, 4169-007 Porto, Portugal}

\author[0000-0001-6396-2563]{Tanda Li}
\affiliation{Department of Astronomy, Beijing Normal University, Beijing, 100875, P.~R.~China}
\affiliation{School of Physics and Astronomy, University of Birmingham, Edgbaston, Birmingham B15 2TT, UK}
\affiliation{Stellar Astrophysics Centre (SAC), Department of Physics and Astronomy, Aarhus University, Ny Munkegade 120, 8000 Aarhus C, Denmark}

\author[0000-0001-7664-648X]{J. M. Joel Ong}
\affiliation{Department of Astronomy, Yale University, 52 Hillhouse Ave., New Haven, CT 06511, USA}
\affiliation{Institute for Astronomy, University of Hawai`i, 2680 Woodlawn Drive, Honolulu, HI 96822, USA}
\affiliation{Hubble Fellow}

\author[0000-0001-8835-2075]{Enrico Corsaro}
\affiliation{INAF --- Osservatorio Astrofisico di Catania, via S.~Sofia 78, 95123 Catania, Italy}

\author[0000-0001-8237-7343]{Margarida S. Cunha}
\affiliation{Instituto de Astrof\'{\i}sica e Ci\^{e}ncias do Espa\c{c}o, Universidade do Porto,  Rua das Estrelas, 4150-762 Porto, Portugal}
\affiliation{Departamento de F\'{\i}sica e Astronomia, Faculdade de Ci\^{e}ncias da Universidade do Porto, Rua do Campo Alegre, s/n, 4169-007 Porto, Portugal}

\author[0000-0001-5222-4661]{Timothy R. Bedding}
\affiliation{Sydney Institute for Astronomy (SIfA), School of Physics, University of Sydney, NSW 2006, Australia}
\affiliation{Stellar Astrophysics Centre (SAC), Department of Physics and Astronomy, Aarhus University, Ny Munkegade 120, 8000 Aarhus C, Denmark}

\author[0000-0002-9480-8400]{Diego Bossini}
\affiliation{Instituto de Astrof\'{\i}sica e Ci\^{e}ncias do Espa\c{c}o, Universidade do Porto, Rua das Estrelas, 4150-762 Porto, Portugal}

\author[0000-0003-0377-0740]{Sylvain N. Breton}
\affiliation{AIM, CEA, CNRS, Universit\'e Paris-Saclay, Universit\'e Paris Diderot, Sorbonne Paris Cit\'e, F-91191 Gif-sur-Yvette, France}

\author[0000-0002-1988-143X]{Derek L. Buzasi}
\affiliation{Department of Chemistry and Physics, Florida Gulf Coast University, 10501 FGCU Blvd.~S., Fort Myers, FL 33965, USA}

\author[0000-0002-5714-8618]{William J. Chaplin}
\affiliation{School of Physics and Astronomy, University of Birmingham, Edgbaston, Birmingham B15 2TT, UK}
\affiliation{Stellar Astrophysics Centre (SAC), Department of Physics and Astronomy, Aarhus University, Ny Munkegade 120, 8000 Aarhus C, Denmark}

\author[0000-0001-6774-3587]{Morgan Deal}
\affiliation{Instituto de Astrof\'{\i}sica e Ci\^{e}ncias do Espa\c{c}o, Universidade do Porto,  Rua das Estrelas, 4150-762 Porto, Portugal}

\author[0000-0002-8854-3776]{Rafael A. Garc\'ia}
\affiliation{AIM, CEA, CNRS, Universit\'e Paris-Saclay, Universit\'e Paris Diderot, Sorbonne Paris Cit\'e, F-91191 Gif-sur-Yvette, France}

\author[0000-0002-0139-4756]{Michelle L. Hill}
\affiliation{Department of Earth and Planetary Sciences, University of California Riverside, 900 University Ave., Riverside, CA 92521, USA}

\author[0000-0003-2400-6960]{Marc Hon}
\affiliation{Institute for Astronomy, University of Hawai`i, 2680 Woodlawn Drive, Honolulu, HI 96822, USA}
\affiliation{School of Physics, The University of New South Wales, Sydney NSW 2052, Australia}

\author[0000-0001-8832-4488]{Daniel Huber}
\affiliation{Institute for Astronomy, University of Hawai`i, 2680 Woodlawn Drive, Honolulu, HI 96822, USA}

\author[0000-0002-7614-1665]{Chen Jiang}
\affiliation{Max-Planck-Institut f\"ur Sonnensystemforschung, Justus-von-Liebig-Weg 3, 37077 G\"ottingen, Germany}

\author[0000-0002-7084-0529]{Stephen R. Kane}
\affiliation{Department of Earth and Planetary Sciences, University of California Riverside, 900 University Ave., Riverside, CA 92521, USA}

\author[0000-0001-9198-2289]{Cenk Kayhan}
\affiliation{Department of Astronomy and Space Sciences, Erciyes University, 38030, Kayseri, Turkey}

\author[0000-0002-3322-5279]{James S. Kuszlewicz}
\affiliation{Center for Astronomy (Landessternwarte), Heidelberg University, K\"onigstuhl 12, 69118 Heidelberg, Germany}

\author[0000-0003-3742-1987]{Jorge Lillo-Box}
\affiliation{Centro de Astrobiolog\'ia (CAB, CSIC-INTA), Depto.~de Astrof\'isica, ESAC Campus 28692 Villanueva de la Ca\~{n}ada (Madrid), Spain}

\author[0000-0002-0129-0316]{Savita Mathur}
\affiliation{Instituto de Astrof\'isica de Canarias (IAC), E-38205 La Laguna, Tenerife, Spain}
\affiliation{Universidad de La Laguna (ULL), Departamento de Astrof\'isica, E-38206 La Laguna, Tenerife, Spain}

\author[0000-0003-0513-8116]{M\'ario J. P. F. G. Monteiro}
\affiliation{Instituto de Astrof\'{\i}sica e Ci\^{e}ncias do Espa\c{c}o, Universidade do Porto,  Rua das Estrelas, 4150-762 Porto, Portugal}
\affiliation{Departamento de F\'{\i}sica e Astronomia, Faculdade de Ci\^{e}ncias da Universidade do Porto, Rua do Campo Alegre, s/n, 4169-007 Porto, Portugal}

\author[0000-0002-2157-7146]{Filipe Pereira}
\affiliation{Instituto de Astrof\'{\i}sica e Ci\^{e}ncias do Espa\c{c}o, Universidade do Porto,  Rua das Estrelas, 4150-762 Porto, Portugal}
\affiliation{Departamento de F\'{\i}sica e Astronomia, Faculdade de Ci\^{e}ncias da Universidade do Porto, Rua do Campo Alegre, s/n, 4169-007 Porto, Portugal}

\author[0000-0003-4422-2919]{Nuno C. Santos}
\affiliation{Instituto de Astrof\'{\i}sica e Ci\^{e}ncias do Espa\c{c}o, Universidade do Porto,  Rua das Estrelas, 4150-762 Porto, Portugal}
\affiliation{Departamento de F\'{\i}sica e Astronomia, Faculdade de Ci\^{e}ncias da Universidade do Porto, Rua do Campo Alegre, s/n, 4169-007 Porto, Portugal}

\author[0000-0001-6359-2769]{Aldo Serenelli}
\affiliation{Institute of Space Sciences (ICE, CSIC) Campus UAB, Carrer de Can Magrans, s/n, E-08193, Bellaterra, Spain}
\affiliation{Institut d'Estudis Espacials de Catalunya (IEEC), Carrer del Gran Capit\`a, 2-4, E-08034, Barcelona, Spain}

\author[0000-0002-4879-3519]{Dennis Stello}
\affiliation{School of Physics, The University of New South Wales, Sydney NSW 2052, Australia}
\affiliation{Stellar Astrophysics Centre (SAC), Department of Physics and Astronomy, Aarhus University, Ny Munkegade 120, 8000 Aarhus C, Denmark}

\begin{abstract}

Asteroseismology is playing an increasingly important role in the characterization of red-giant host stars and their planetary systems. Here, we conduct detailed asteroseismic modeling of the evolved red-giant branch (RGB) hosts KOI-3886 and $\iota$ Draconis, making use of end-of-mission \kepler\ (KOI-3886) and multi-sector TESS ($\iota$ Draconis) time-series photometry. We also model the benchmark star KIC~8410637, a member of an eclipsing binary, thus providing a direct test to the seismic determination. We test the impact of adopting different sets of observed modes as seismic constraints. Inclusion of $\ell=1$ and 2 modes improves the precision on the stellar parameters, albeit marginally, compared to adopting radial modes alone, with $1.9$--$3.0\%$ (radius), $5$--$9\%$ (mass), and $19$--$25\%$ (age) reached when using all p-dominated modes as constraints. Given the very small spacing of adjacent dipole mixed modes in evolved RGB stars, the sparse set of observed g-dominated modes is not able to provide extra constraints, further leading to highly multimodal posteriors. Access to multi-year time-series photometry does not improve matters, with detailed modeling of evolved RGB stars based on (lower-resolution) TESS data sets attaining a precision commensurate with that based on end-of-mission \kepler\ data. Furthermore, we test the impact of varying the atmospheric boundary condition in our stellar models. We find mass and radius estimates to be insensitive to the description of the near-surface layers, at the expense of substantially changing both the near-surface structure of the best-fitting models and the values of associated parameters like the initial helium abundance, $Y_{\rm i}$. Attempts to measure $Y_{\rm i}$ from seismic modeling of red giants may thus be systematically dependent on the choice of atmospheric physics.

\end{abstract}

\keywords{asteroseismology --- stars: evolution --- stars: fundamental parameters --- stars: individual (HD~190655, HD~137759, TYC~3130-2385-1)}

\section{Introduction} \label{sec:intro}

Throughout the course of the NASA \kepler\//K2 mission \citep{Borucki10,Howell14}, asteroseismology has played an important role in the characterization of host stars and their planetary systems \citep[for recent reviews, see][]{CampanteBook,Lundkvist18}. \kepler\//K2 mainly targeted main-sequence stars, however, preventing a systematic transit survey of red giants and hence robust inference of the planet occurrence around such stars \citep{Huber13,Box14,Grunblatt17,Grunblatt19}. This meant that the synergy between asteroseismology and exoplanetary science would remain mostly confined to unevolved stars. The advent of NASA's \textit{Transiting Exoplanet Survey Satellite} \citep[TESS;][]{TESS} has since lifted this restriction, raising the yield of oscillating giants to a few hundreds of thousands \citep{Hon21}, an order of magnitude increase over the yield from \kepler\ and K2. This has enabled the systematic search for transiting planets around seismic giants \citep{Campante16,Huber19,Pereira19,Grunblatt22,Grunblatt22-iii,Saunders22}, as well as revisiting previously known (mostly from radial-velocity surveys) evolved hosts using asteroseismology \citep{Campante19,Ball20,Jiang20,Jiang23,Malla20,Nielsen20,Kane21}.

\begin{figure}[!t]
	\centering
	\includegraphics[width=.48\textwidth,trim=0.5cm 0.2cm 0.2cm 0cm,clip]{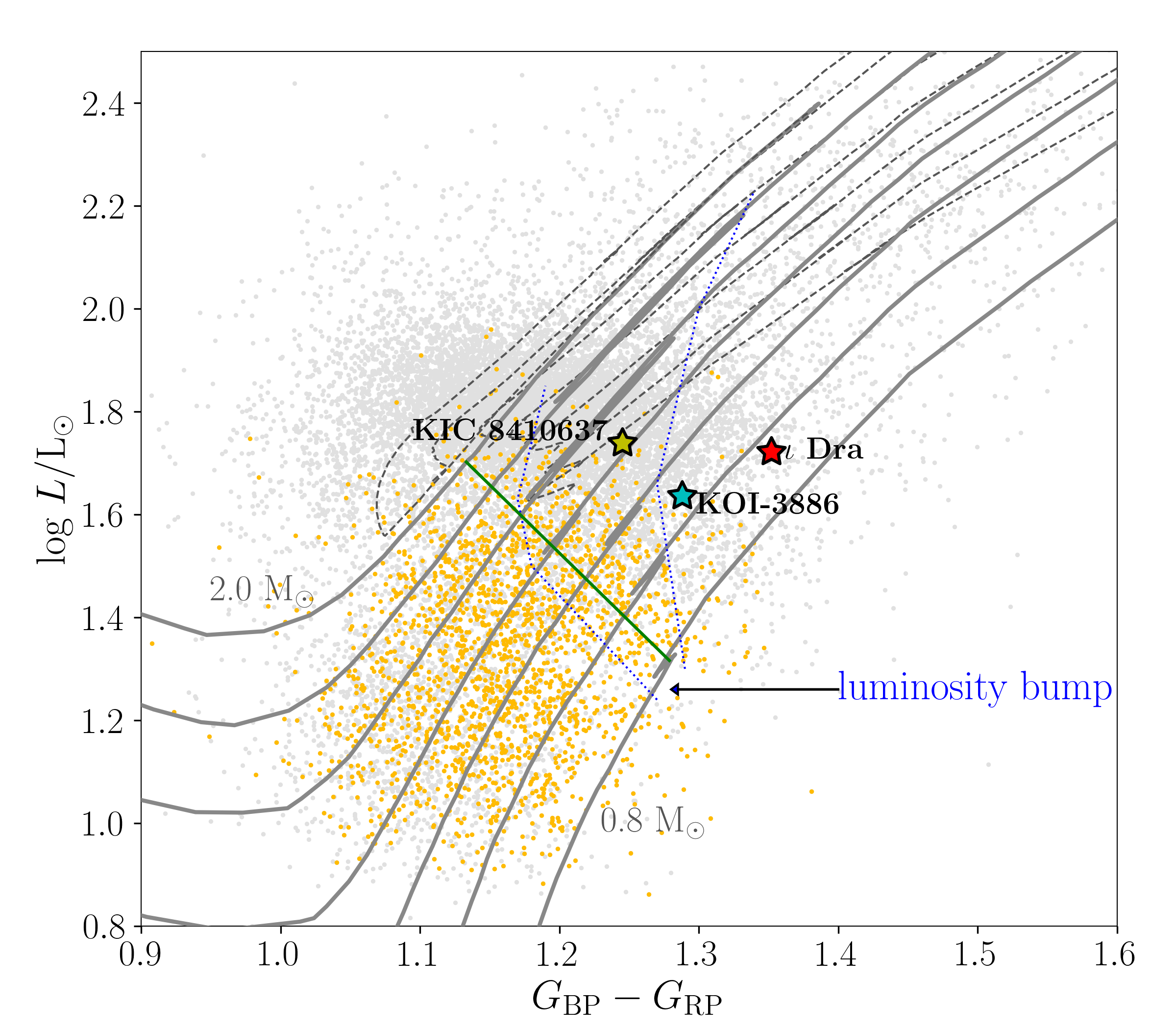}
	\caption{Location of KOI-3886 and $\iota$ Dra in an HR diagram, where the (dereddened) \gaia\ $G_{\rm BP}-G_{\rm RP}$ color index is used as a proxy for \teff. The benchmark star KIC~8410637 (see Sect.~\ref{sec:modeling}) is also displayed. Gray dots represent the $\sim 16{,}000$ \kepler\ seismic red giants from the catalog of \citet{Yu18}. Orange dots highlight the subset of RGB stars with a measured gravity-mode period spacing using the automated approach of \citet{Vrard16}, with most such stars restricted to luminosities lower than that of the red-giant luminosity bump. Stellar evolutionary tracks \citep[][]{PARSEC}, ranging in mass from $0.8$ to $2.0\,{\rm M}_\odot$ (in steps of $0.2\,{\rm M}_\odot$), are shown as gray solid curves while on the RGB (the solid green line connects models with $\Delta\nu \sim 6\:{\rm \mu Hz}$). Models in the tracks have ${\rm \metal}=-0.1\:{\rm dex}$, typical of stars in the Yu et al.~catalog (note that the three stars studied herein all have supersolar metallicities instead; see Table \ref{tab:starparam}). The locus of the luminosity bump (across tracks) is delimited by the curved blue dotted lines.}
	\label{fig:HRD}
\end{figure}

Here, we revisit the evolved\footnote{We herein adopt the term \emph{evolved RGB} to denote stars beyond the limit of visibility of gravity-dominated mixed modes on the RGB, i.e., with $\Delta\nu \lesssim 6\:{\rm \mu Hz}$ \citep[][]{Mosser18}. See Sect.~\ref{sec:seismology} for a definition of $\Delta\nu$.} RGB host stars KOI-3886~A (HD~190655, KIC~8848288, TIC~185060864; hereafter KOI-3886) and $\iota$ Draconis (HD~137759, TIC~165722603; hereafter $\iota$ Dra), making use of end-of-mission \kepler\ (KOI-3886) and multi-sector TESS ($\iota$ Dra) time-series photometry. Their properties, as found in the literature, are compiled in Table \ref{tab:starparam}. Figure \ref{fig:HRD} shows their location in a Hertzsprung--Russell (HR) diagram. KOI-3886, observed continuously by \kepler\ for nearly 4 years, has been a longtime candidate host \citep{Rowe15}. In \citet{Lillo-Box21}, we concluded that the close-in planet candidate is in fact a false positive and reinterpreted the system as an eclipsing brown dwarf in a hierarchical triple containing two evolved stars. The fundamental stellar parameters derived from asteroseismology for KOI-3886 (the primary) were central to that study, entering an iterative procedure to determine the final set of parameters for the three bodies in the system.

$\iota$ Dra, known for two decades to host a planet\footnote{$\iota$ Dra b was the first planet found to orbit a giant star \citep{Frink02}.} in a highly eccentric, 511-day-period orbit \citep{Frink02,Zechmeister08,Kane10}, was observed by TESS over 5 noncontiguous sectors (each sector is 27.4 days long) during the second year of its nominal mission. In \citet{Hill21}, we presented the results of continued radial-velocity (RV) monitoring of $\iota$ Dra over several orbits of its known planet. The newly acquired RV observations allowed detecting curvature in the previously identified RV trend, which was interpreted as likely being caused by an outer companion. Through the combination of the RV measurements with space astrometry, we confirmed the presence of an additional long-period, eccentric companion. Mass predictions from our analysis --- which used the seismic mass derived for $\iota$ Dra as a prior --- place the companion on the border of the planet and brown dwarf regimes.

The presence of planets orbiting KOI-3886 and $\iota$ Dra (putative in the former case) is what originally prompted the seismic analyses --- conducted separately, although not in a strictly independent manner --- of these two red-giant stars. In this follow-up work, we give a full account of the seismic analyses underpinning \citet{Lillo-Box21} and \citet{Hill21}, while making an incursion into the wider problematic of detailed asteroseismic modeling of red giants. Evolutionary calculations for RGB stars are especially sensitive to small variations in the choices of input physics and model parameters in a highly nonlinear fashion. For instance, small variations in the amount of envelope overshooting result in shifts to both the age and luminosity at which the red-giant luminosity bump occurs, as well as lateral adjustments to the position of the RGB itself in the HR diagram \cite[e.g.,][]{Khan18}. These changes are degenerate with those induced by variations in the initial helium abundance, $Y_{\rm i}$, and the mixing-length parameter, $\alpha_\mathrm{MLT}$, neither of which can be directly constrained in cool stars. The same is also true of other inputs to stellar evolution, such as the reference values used for solar composition, equations of state, or opacity tables, which are not typically treated as variable parameters in grids of stellar models used for this purpose.

To make matters more complicated, observed oscillation modes for red giants are of mixed g- (or gravity) and p-mode (or pressure) nature, resulting from the coupling between buoyancy waves that propagate throughout the core and acoustic waves propagating in the stellar envelope \cite[e.g.,][]{Aizenman77,UnnoBook}. As a star evolves along the RGB, its g- and p-mode cavities become increasingly decoupled, resulting in surface amplitudes for the g-dominated mixed modes that are too low to make such modes readily detectable, even if allowing enough time to fully resolve them \citep[][]{Grosjean14,Mosser18}. Consequently, one ends up with a paucity of observed g-dominated modes per radial order, making the characterization of the mixed-mode pattern for evolved RGB stars potentially challenging \citep[see Fig.~\ref{fig:HRD};][]{Stello13,Mosser15,Mosser19,Vrard16,HAYDN}.

The aim of this work is thus threefold. First, we provide consolidated parameters for the evolved RGB hosts KOI-3886 and $\iota$ Dra through detailed asteroseismic modeling (Sects.~\ref{sec:modeling_opt}--\ref{sec:interdisp}). To that end, we carry out a comprehensive stellar characterization by employing two independent and well-established modeling pipelines \citep{2020MNRAS.495.3431L,Ong21}, hence further allowing the (inter-pipeline) systematics on the inferred stellar parameters --- arising from the use of different model grids, input physics, and analysis methodologies --- to be estimated. Critically, we include methodological documentation and discussion that had been deferred to this work from \citet{Lillo-Box21} and \citet{Hill21}, where results from a single modeling pipeline \citep[namely, that of][]{2020MNRAS.495.3431L} were only briefly presented.

In addition to the above two stars, we also model the \kepler\ benchmark star KIC~8410637 \citep[TIC~123417372, TYC~3130-2385-1;][]{2013A&A...556A.138F,Gaulme16,Li18,Li22,Themessl18}, a member of an eclipsing binary, thereby providing a direct test to the seismic determination. All three stars in our sample are of relatively low mass (i.e., $M\lesssim1.8\,{\rm M}_\odot$, thus eventually igniting helium in the core in degenerate conditions), in accordance with \kepler\ observations of seismic evolved RGB stars \citep[see Fig.~\ref{fig:HRD}; e.g.,][]{Mosser12,Mosser14,Stello13,Vrard16}.

Second, we test the impact of the optimization procedure on the inferred stellar parameters of evolved RGB stars by basing the detailed asteroseismic modeling on three alternative, nested sets of seismic constraints (Sect.~\ref{sec:modeling_opt}), namely, by using radial modes alone, by including all p-like modes, and ultimately by adopting the full mode frequency lists (which include g-dominated modes). Compared to the conventional use of asteroseismic scaling relations, detailed modeling using radial modes has been shown to significantly improve on the accuracy of radius and mass estimates for RGB stars, while reaching a typical (median) precision of $1.7\%$, $4.5\%$, and $16\%$ respectively on the radius, mass, and age \citep{Li22}, reasons that motivated our decision to conduct detailed modeling to begin with.

The modeling of individual mode frequencies --- other than radial modes --- of red giants in general is, however, still in a rudimentary stage. This can be attributed to the computational expense involved in the numerical evaluation of mixed modes \citep{Li18,Ong21b}, with evolved RGB stars in particular having very densely spaced (theoretical) dipole mixed modes, as well as to the much needed improvements in the modeling of red giants so as to fully realize the observational accuracy of mode frequencies, particularly that of nonradial modes \citep{Aarhus2,Aarhus1}. Notwithstanding, we make use of our sample of three seismic evolved RGB stars, for which high-quality, multi-year \kepler\ (KOI-3886 and KIC~8410637) and multi-sector TESS ($\iota$ Dra) time-series photometry is available, to gain insight into the constraining power of nonradial modes (including $\ell = 1$ g-dominated modes) when applied to the detailed modeling of this specific type of star.

Finally, we test the impact of the choice of near-surface physics, namely, by varying the atmospheric boundary condition (Sect.~\ref{sec:modeling_phys}). Limitations in the numerical modeling of the near-surface layers of stars induce errors on the mode frequencies computed from stellar models, the so-called asteroseismic \emph{surface term}. This surface term presents a significant methodological obstacle to the use of individual p-mode frequencies in seismic modeling in general. The determination of methods by which it may be corrected for remains an area of active research \citep[e.g.,][]{Ball14,Compton18,Nsamba18,Yaguang22}. A correction for the surface term in red giants requires special treatment, owing to the presence of mixed modes \citep{Ball18,Ong21b}, and estimates of some of the properties of red giants have been shown to be potentially sensitive to methodological decisions as to how this correction is to be performed \citep[e.g.,][]{Ong21}.

One operational assumption underlying such studies has, however, been less well examined. Specifically, it is assumed that these near-surface modeling errors change only the seismic properties of the star, and either do not significantly modify the spectroscopic surface observables, or modify them in a manner that can be calibrated away by the appropriate choice of other tuning parameters, such as the convective mixing length. While this may be a suitable approximation for the study of main-sequence stars, the locus of the RGB itself depends sensitively on the choice of the atmospheric boundary condition used in numerical stellar evolution, and so in these cases this approximation no longer holds. The interplay between these changes to the spectroscopic surface properties and the effects of the surface-term correction may result in further methodological dependences for stellar parameters --- like radii, masses, and ages --- derived with stellar modeling and asteroseismology. Since this sensitivity to inputs of numerical stellar evolution increases with luminosity, we examine how these confounding effects affect evolved RGB stars in particular, for which we expect any such systematic issues to be most significant.

The rest of this paper is organized as follows. In Sect.~\ref{sec:phot}, we present the adopted \kepler\ and TESS photometry. This is followed in Sect.~\ref{sec:seismology} by the analysis of both data sets, including the measurement of individual mode frequencies. Detailed asteroseismic modeling is performed in Sect.~\ref{sec:modeling}. Finally, a summary and conclusions are presented in Sect.~\ref{sec:conclusions}.

\begin{table*}[!t]
\begin{center}
\caption{Stellar parameters for KOI-3886, $\iota$ Dra, and KIC~8410637.\label{tab:starparam}}
\renewcommand{\tabcolsep}{1mm}
\begin{tabular}{l | c c | c c | c c}
\noalign{\smallskip}
\tableline\tableline
\noalign{\smallskip}
\multicolumn{1}{l}{} & \multicolumn{2}{c}{\textbf{KOI-3886}} & \multicolumn{2}{c}{\textbf{$\iota$ Dra}} & \multicolumn{2}{c}{\textbf{KIC~8410637}\tablenotemark{a}} \\
\noalign{\smallskip}
\tableline
\noalign{\smallskip}
Parameter & Value & Source & Value & Source & Value & Source \\
\noalign{\smallskip}
\hline
\noalign{\smallskip}
\multicolumn{7}{c}{\gaia\ Photometry and Parallax} \\
\noalign{\smallskip}
\hline
\noalign{\smallskip}
DR3 ID & 2082133182277361152 & 1 & 1614731957531452544 & 1 & 2105415749007167616 & 1 \\
$G$-band Mag. & 10.1 & 1 & 2.97 & 1 & 10.8 & 1 \\
$G_{\rm BP}-G_{\rm RP}$ & 1.29 & 1 & 1.35 & 1 & 1.24 & 1 \\
$\pi$ (mas) & 2.139$\pm$0.307 & 1 & 32.52$\pm$0.14 & 1 & 0.839$\pm$0.018 & 1 \\
\noalign{\smallskip}
\hline
\noalign{\smallskip}
\multicolumn{7}{c}{Spectroscopy} \\
\noalign{\smallskip}
\hline
\noalign{\smallskip}
$T_{\rm eff}$ (K) & 4720$\pm$120 & 2 &4504$\pm$62 & 3,4 &4750$\pm$86 & 5 \\
$[{\rm Fe}/{\rm H}]$ (dex) & 0.14$\pm$0.07 & 2 &0.03$\pm$0.08 & 3,4 & 0.12$\pm$0.08 & 5 \\
$\log g$ (cgs) & 2.54$\pm$0.24& 2 & 2.52$\pm$0.07 & 3,4 &2.75$\pm$0.15 & 5 \\
\noalign{\smallskip}
\hline
\noalign{\smallskip}
\multicolumn{7}{c}{{\it Reference} Fundamental Stellar Parameters} \\
\noalign{\smallskip}
\hline
\noalign{\smallskip}
$M$ (${\rm M}_\odot$) & \nodata & \nodata & \nodata & \nodata & 1.56$\pm$0.03\tablenotemark{b} & 6 \\
$R$ (${\rm R}_\odot$) & \nodata & \nodata & 11.99$\pm$0.06\tablenotemark{c} & 7 & 10.74$\pm$0.11\tablenotemark{b} & 6 \\
$L$ (${\rm L}_\odot$) & 43.3$\pm$9.5\tablenotemark{d} & 2 & 52.8$\pm$2.1\tablenotemark{d} & 4 & 54.8$^{+4.0}_{-3.7}$\tablenotemark{e} & 6 \\
\noalign{\smallskip}
\hline
\noalign{\smallskip}
\multicolumn{7}{c}{Global Oscillation Parameters} \\
\noalign{\smallskip}
\hline
\noalign{\smallskip}
$\Delta\nu$ ($\mu$Hz) & $4.60 \pm 0.20$ & 8 & $4.02 \pm 0.02$ & 8 & 4.63$\pm$0.01 & 5 \\
$\nu_{\rm max}$ ($\mu$Hz) & $46.9 \pm 0.3$ & 8 & $38.4 \pm 0.5$ & 8 & 46.3$\pm$0.9 & 5 \\
\noalign{\smallskip}
\hline
\noalign{\smallskip}
\end{tabular}
\end{center}
\tablenotetext{a}{\scriptsize Benchmark star (see Sect.~\ref{sec:modeling}).}
\tablenotetext{b}{\scriptsize From the dynamical modeling of the eclipsing binary's orbit.}
\tablenotetext{c}{\scriptsize From interferometry.}
\tablenotetext{d}{\scriptsize From a fit to the spectral energy distribution (SED).}
\tablenotetext{e}{\scriptsize From a joint light and velocity curve analysis.}
\tablerefs{\scriptsize (1) \citet{Gaia}, (2) \citet{Lillo-Box21}, (3) \citet{Jofre15}, (4) \citet{Hill21}, (5) \citet{Li18}, (6) \citet{2013A&A...556A.138F}, (7) \citet{Baines11}, (8) this work.}
\end{table*}

\section{Photometry} \label{sec:phot}

KOI-3886 was observed by \kepler\ in long-cadence mode (29.4 minutes) between Quarters 0 and 17 (or continuously for nearly 4 years). It was since also observed by TESS during Sectors 14--15, 41, and 54--56. Owing to the lower precision and relatively shorter temporal coverage of the TESS observations, we have nonetheless decided to base the seismic analysis of KOI-3886 solely on the available \kepler\ photometry. We make use of a {\sc kepseismic}\footnote{\url{https://archive.stsci.edu/prepds/kepseismic/}} light curve, which has been optimized for asteroseismology. The light curve was extracted from the target pixel files using a custom aperture and subsequently processed through the {\sc kadacs} pipeline \citep[\kepler\ Asteroseismic Data Analysis and Calibration Software;][]{KADACS1}. {\sc kadacs} corrects for outliers, jumps, and drifts, also filling any gaps shorter than 20 days using in-painting techniques \citep{KADACS2,KADACS3}. The light curve was further high-pass filtered using an 80-day triangular smoothing function, after which it underwent an iterative (three iterations) $\sigma$-clipping procedure (3$\sigma$ level) in order to mitigate the impact of the eclipses on the computation of the power spectrum.

$\iota$ Dra was observed by TESS over 5 noncontiguous sectors (namely, Sectors 15--16 and 22--24) at a 2-minute cadence during the second year of its nominal mission. With an apparent TESS magnitude of $T=2.27$, $\iota$ Dra is significantly saturated in the TESS photometry. To deal with the target's saturated nature, a large custom aperture was adopted and a background model applied to account for the spatially-varying background light. Full details on the light curve preparation are presented in \citet{Hill21}.

\subsection{On the Potential Transit of $\iota$ Dra b} \label{sec:transit_iotaDrab}

$\iota$~Dra was since again observed by TESS during Sectors 49--51 as part of the extended mission. As noted in \citet{Hill21}, Sector 50 coincided with the expected time of conjunction for $\iota$~Dra~b. Using the mass-radius probabilistic modeling tool {\sc forecaster} \citep{chen2017}, and adopting the planet's minimum mass \citep[$M_{\rm p}\sin i = 11.82^{+0.42}_{-0.41}\:M_{\rm Jup}$;][]{Hill21}, we estimate its radius to be $1.10^{+0.22}_{-0.19}\:R_{\rm Jup}$. This gives an expected transit signal\footnote{The transit probability of $\iota$~Dra~b is of $\sim 16\%$ \citep{Kane10}.} on the large, $11.99\:{\rm R}_\odot$ star of $0.007\%$. After careful analysis, we found no significant transit signature within the light curve. A transit cannot, however, be completely ruled out due to the difficulty in divorcing the predicted small transit signal from the intrinsic stellar variability \citep[e.g.,][]{Pereira19}.

\section{Asteroseismology} \label{sec:seismology}

\subsection{Global Oscillation Parameters} \label{sec:global_osc}

\begin{figure*}[!t]
\centering
  \subfigure{%
  \includegraphics[width=.48\textwidth]{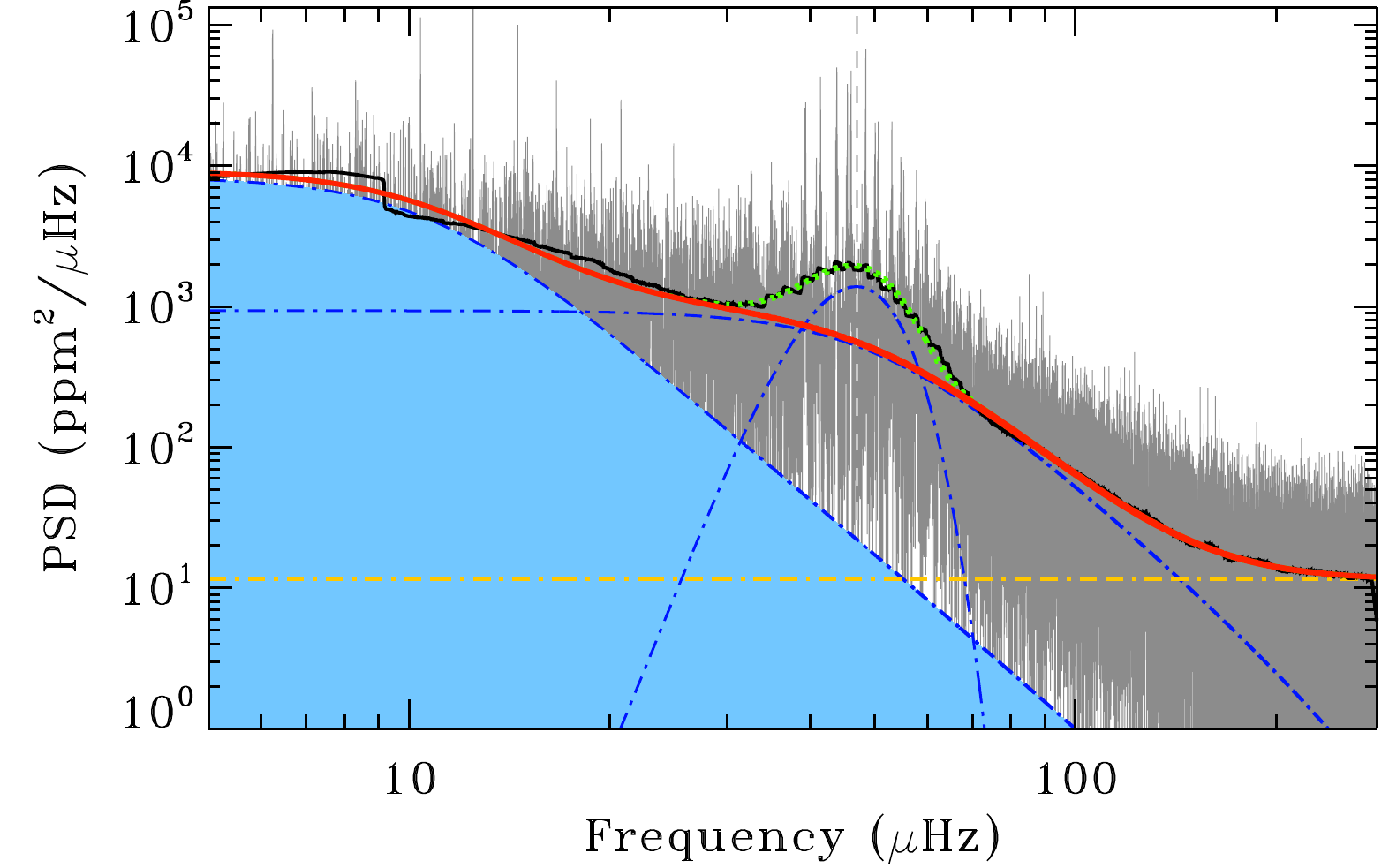}}\hfill
  \subfigure{%
  \includegraphics[width=.48\textwidth]{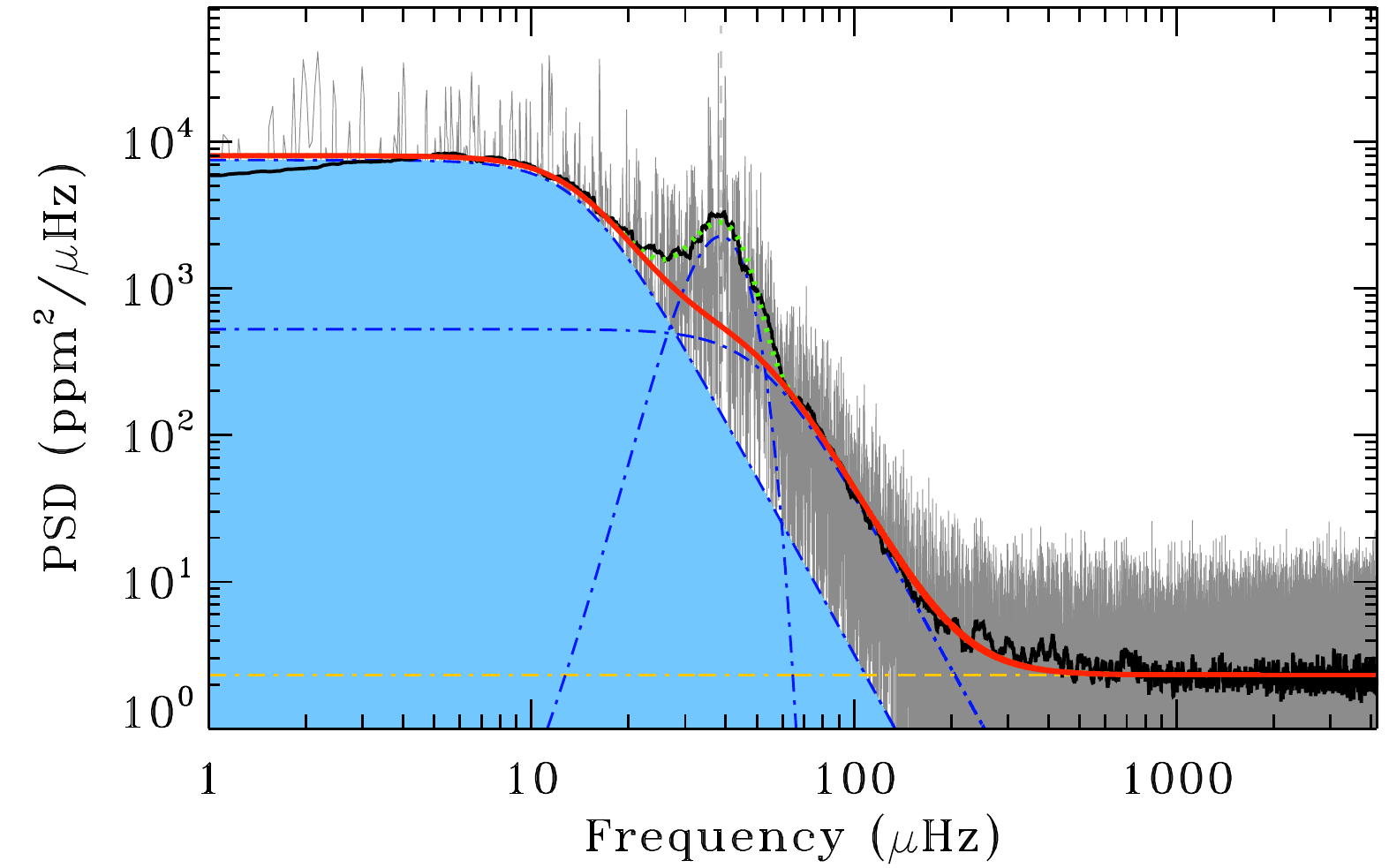}}
  \caption{Power spectral density (PSD) of KOI-3886 (left panel) and $\iota$ Dra (right panel). Power spectra are shown in gray (heavily smoothed version in black). Vertical dashed lines represent \numax. Solid red curves are fits to the background performed with {\sc diamonds} \citep{DIAMONDS}, consisting of two Harvey-like profiles (blue dot-dashed curves) and a white noise offset (yellow dot-dashed line). Joint fits to the oscillation power excess (blue dot-dashed Gaussian) and background are visible as green dotted curves near \numax. Note the smaller frequency range of the PSD of KOI-3886, owing to the lower Nyquist frequency of \kepler's long-cadence data. A residual signature of the eclipse harmonics can still be seen at the low-frequency end in the PSD of KOI-3886.}\label{fig:bckg}
\end{figure*}

Figure \ref{fig:bckg} shows the power density spectra of KOI-3886 (left panel) and $\iota$ Dra (right panel) based on the Lomb--Scargle periodogram \citep{Lomb76,Scargle82} of the light curves extracted in Sect.~\ref{sec:phot}. These reveal a clear power excess due to solar-like oscillations at $\sim50\:{\rm \mu Hz}$ and $\sim40\:{\rm \mu Hz}$, respectively. We measured the large frequency separation, \dnu, and the frequency of maximum oscillation amplitude, \numax, of both stars using a range of well-tested automated methods \citep[e.g.,][]{Huber09,Mathur10,DIAMONDS,Campante17,Campante19,FAMED}. Consolidated pairs of values are listed in Table \ref{tab:starparam} and stem from a single method for each star \citep[for details, see][]{Lillo-Box21,Hill21}, their uncertainties being the corresponding formal uncertainties.

\subsection{Individual Mode Frequencies} \label{sec:peak_bagging}

We searched for individual mode frequencies with angular degree up to $\ell=2$ in the power spectrum of each star (a process dubbed \emph{peak-bagging}) using the {\sc famed} pipeline \citep[Fast and AutoMated pEak bagging with {\sc diamonds};][]{FAMED}. As no definite detection of the presence of rotational splittings could be initially established for these evolved RGB stars \citep[cf.][]{Mosser_corespindown,Gehan2018}, a model for the power spectrum that does not include the effect of rotation was eventually used. Tables \ref{tab:freq_KOI-3886} and \ref{tab:freq_iotaDra} list all significant modes\footnote{The Doppler shift of the observed mode frequencies due to the line-of-sight motion \citep{Davies14} is not significant for both stars and hence no correction has been applied.} returned by {\sc famed} for KOI-3886 and $\iota$ Dra, respectively. A total of 31 (23) modes were extracted across 8 (7) radial orders for KOI-3886 ($\iota$ Dra). Figures \ref{fig:freq_KOI-3886} and \ref{fig:freq_iotaDra} illustrate the outcome of the peak-bagging process using {\sc famed}.

Owing to the lower resolution of the TESS power spectrum of $\iota$ Dra, we have introduced an additional step for selecting mode frequency lists for this star, which combines the output of several peak-bagging procedures. This was mainly motivated by the need to robustly identify and measure long-lived mixed modes. Several \emph{peak-baggers} ($N=7$) initially extracted individual mode frequencies from the power spectrum of $\iota$ Dra. The methods employed included both iterative sine-wave fitting \citep[e.g.,][]{PERIOD04} as well as the fitting of Lorentzian and sinc$^2$ mode profiles \citep[e.g.,][]{HandCamp}, the latter being the approach implemented in {\sc famed}. Two frequency lists were then produced following the procedure described in \citet{Campante11}, namely, a \emph{maximal frequency list} and a \emph{minimal frequency list}. The former includes modes detected by at least 2 peak-baggers, whereas the latter includes only those modes detected by more than $\lfloor N/2\rfloor$ peak-baggers. The more conservative minimal list is thus a subset of the maximal list. Importantly, modes in the minimal list are the ones subject to detailed modeling in Sect.~\ref{sec:modeling}. To guarantee reproducibility, we resort to a set of observed mode frequencies (and corresponding uncertainties) tracing back to a single method ({\sc famed}), as opposed to an averaged set. We note that \citet{Zechmeister08}, using RV measurements, detected the presence of solar-like oscillations in $\iota$ Dra with frequencies around $3$--$4\:{\rm d^{-1}}$ ($\sim34.7$--$46.3\:{\rm \mu Hz}$), fully consistent with our results. The dominant mode found by those authors ($3.45\:{\rm d^{-1}}$ or $\sim39.9\:{\rm \mu Hz}$) coincides with one of the radial modes in the minimal list (see Table \ref{tab:freq_iotaDra}).

\subsection{Evolutionary State} \label{sec:evol_state}
Prior knowledge of the evolutionary state of both stars is crucial towards an accurate determination of their fundamental parameters in Sect.~\ref{sec:modeling}. KOI-3886 has been classified in the literature as a hydrogen-shell burning red giant following a number of complementary analyses of its oscillation power spectrum, namely, based on the pressure-mode pattern \citep{Kallinger12,Vrard18}, the morphology of the mixed modes \citep{Elsworth17}, and deep learning \citep{Hon17}. Moreover, the observed (pairwise) period spacing can be estimated for the highest-frequency radial order ($\Delta P \sim 50 \: {\rm s}$), which supports this classification \citep[][]{Bedding11,Stello13}. Despite this, the asymptotic period spacing, $\Delta\Pi_1$, could not be reliably inferred due to the paucity of observed g-dominated dipole mixed modes.

\begin{figure}[!t]
	\centering
	\includegraphics[width=.46\textwidth,trim=0.2cm 0.2cm 0.2cm 0.0cm,clip]{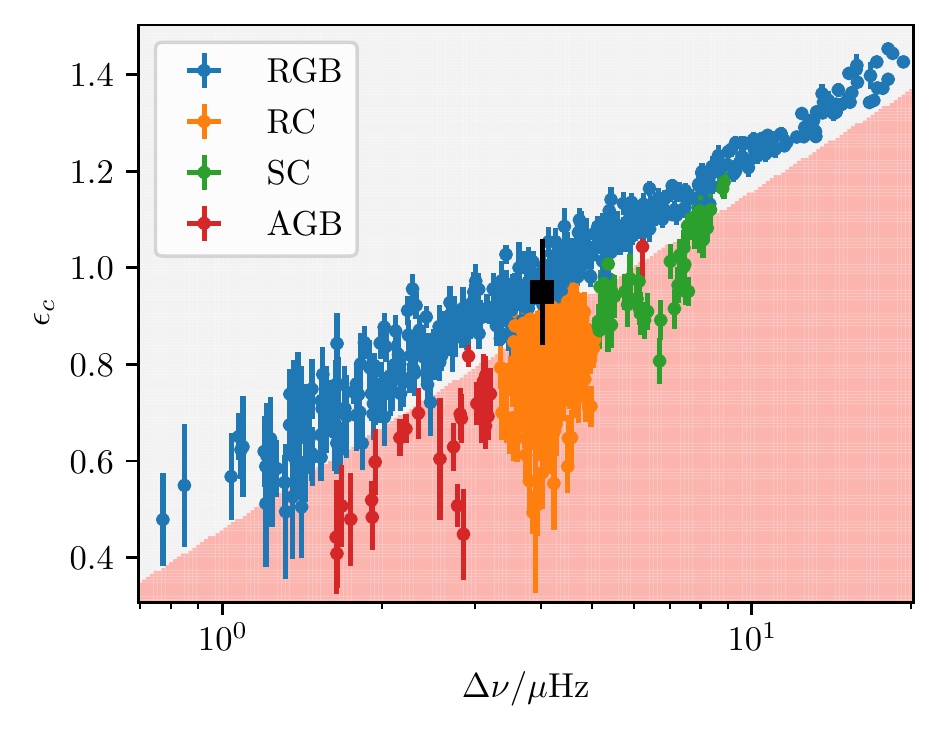}
	\caption{Classification of $\iota$ Dra using the p-mode phase offset, $\varepsilon_{\rm c}$. We show the data set from fig.~4 of \cite{Kallinger12} as colored points (`RGB' = red-giant branch, `RC' = red clump, `SC' = secondary clump, `AGB' = asymptotic-giant branch) and the measured value of $\varepsilon_{\rm c}$ for $\iota$ Dra as the black square. In the background, we show, using shaded regions, the decision boundary of a one-vs-rest support vector machine classifier fitted against the first-ascent red giants of \cite{Kallinger12}. $\iota$ Dra lies close enough to the decision boundary that either a first-ascent red giant or red-clump star are plausible descriptions of it.}
	\label{fig:epsilon}
\end{figure}

The evolutionary state of $\iota$ Dra, on the other hand, remains (seismically) unclassified, and we have adopted a number of approaches in this work to address this. Given the limited number of observed dipole mixed modes per radial order, estimation of $\Delta P$ based on modes in the minimal list proved unsuccessful. An attempt was then made at constraining $\Delta\Pi_1$ based on the stretching of the power spectrum \citep{Vrard16}, which again was inconclusive. Alternatively, the asymptotic acoustic-mode offset, $\varepsilon_{\rm c}$, can in principle be used as a discriminant between hydrogen-shell burning (RGB) and helium-core burning (HeB) red giants \citep{Kallinger12,JCD14}. The measured value of $\varepsilon_{\rm c}=0.95\pm0.11$, however, lies very close (within $1\sigma$) to the decision boundary of \cite{Kallinger12}, potentially allowing for either evolutionary state with this rudimentary method (see \cref{fig:epsilon}).

Machine learning classification methods provided the first robust indication that $\iota$ Dra is an RGB star. We ran the time-domain classifier {\sc Clumpiness} \citep{Clumpiness} over the full TESS light curve, as well as the two contiguous data subsets, i.e., Sectors 15--16 and Sectors 22--24. The probability of the star being on the RGB is respectively $p=0.78$, 0.56, and 0.81. The corresponding probability of it being an HeB star is $p=0.21$, 0.44, and 0.19, with the remaining (negligible) probability being assigned to a \emph{noise} class (e.g., main sequence). Moreover, application of the deep learning classification method of \citet{Hon17,Hon18} to the full TESS light curve, which uses folded background-corrected power spectra and \dnu\ as input, gives support to this by favoring an RGB scenario with high confidence ($p>0.9$).

Finally, we assessed the evolutionary state of $\iota$ Dra based on a preliminary grid-based modeling exercise, having considered as observational constraints its global oscillation parameters, \teff, \metal, and luminosity (see Table \ref{tab:starparam}). We modeled the star twice, assuming it is either on the RGB or in the red clump (RC). We found that assuming the star to be on the RGB yields a solution fully consistent with the observational constraints as determined by comparing their posterior and input values. On the contrary, if the RC evolutionary state is assumed, the grid-based modeling yields a solution for which the posterior values of the combination \teff--\metal\ are inconsistent with the input data at the $4\sigma$ level, i.e., RC model tracks are unable to simultaneously reproduce the \teff\ and \metal\ constraints. In particular, this solution would require a metallicity higher than that observed by at least 0.3 dex. As a result, the total probability of the star being on the RGB is several orders of magnitude higher than it being in the RC. Interpreting this in terms of a Bayes' factor provides decisive evidence in favor of the RGB scenario given the adopted set of observational constraints.

Moreover, if the luminosity constraint is dropped altogether, the RGB solution leads to $L=55.3\pm2.8\:{\rm L}_\odot$, in excellent agreement with the reference value. In contrast, the RC solution then yields $L=69.5\pm3.3\:{\rm L}_\odot$, which is inconsistent with the reference value at the $3\sigma$ level. In this case, the only way to reconcile matters would be to induce a global shift in the \teff\ scale of RC models (and RGB models alike) so that they become cooler by about $300\:{\rm K}$. This large shift is, however, not supported by any previous work on seismic giants. We have nonetheless tested different \teff\ scales by employing models with $\alpha_\mathrm{MLT}$ ranging from 2.1 (solar-calibrated value with a Krishna Swamy atmosphere) down to 1.8 (solar-calibrated value with an Eddington atmosphere). As expected, the coolest models (Eddington atmosphere) lead to a lower luminosity, $L=61.2\pm3.9\:{\rm L}_\odot$, although still in disagreement with the reference value. We note that the RGB solution continues to point to a value, $L=53.4\pm3.8\:{\rm L}_\odot$, in excellent agreement with the reference one. The total probability of the RC solution is lower than that of the RGB case by more than three orders of magnitude. We have thus confirmed, for a reasonable shift in the \teff\ scales of RGB and RC models, our conclusion that the RGB evolutionary state is strongly favored.

As a final step, we tested (both with and without a luminosity constraint) whether this conclusion is robust against the adopted spectroscopic constraints \citep[cf.][]{Campante19}, by considering the \teff\ and \metal\ derived for this star from high-resolution optical and near-infrared CARMENES spectra \citep[see table A2 of][]{Marfil20}. Despite the significantly higher CARMENES \teff\ ($4836\pm87\:{\rm K}$ versus $4504\pm62\:{\rm K}$ adopted herein), similar considerations to the ones above can be made.

\section{Detailed Modeling} \label{sec:modeling}

This work makes use of two independent and well-established pipelines --- hereafter labeled `TL' \citep{2020MNRAS.495.3431L} and `JO' \citep{Ong21} --- for the detailed asteroseismic modeling. We resort to the TL Pipeline for testing the impact of the optimization procedure on the inferred stellar parameters of evolved RGB stars (Sect.~\ref{sec:modeling_opt}), investigating how different sets of observed oscillation modes contribute to the characterization of this specific type of star. We next use the JO Pipeline to assess how the choice of near-surface physics, more specifically, of the atmospheric boundary condition, modifies the inferred stellar parameters (Sect.~\ref{sec:modeling_phys}). Since the two pipelines employ different underlying grids of stellar models and analysis methodologies, we are able to roughly characterize the relative importance of the above methodological decisions (Sect.~\ref{sec:interdisp}).

Our sample consists of three typical seismic evolved RGB stars, in the sense that they are all characterized by a paucity of observed g-dominated dipole mixed modes, being further of relatively low mass (i.e., $M\lesssim1.8\,{\rm M}_\odot$). Besides the host stars KOI-3886 and $\iota$ Dra, we also model the benchmark star KIC~8410637, for which multi-year \kepler\ time-series photometry is available. This star was selected both for being in an (detached) eclipsing binary, as well as for having a \numax\ similar to that of KOI-3886 and $\iota$ Dra. Owing to the former attribute, its mass and radius have been accurately determined via the dynamical modeling of the eclipsing binary’s orbit \citep{2013A&A...556A.138F}, and can thus provide a direct test to the seismic determination. Observed mode frequencies for KIC~8410637 are taken from table A2 of \citet{Li18}. We also note that a precise ($0.5\%$) interferometric radius is available for $\iota$ Dra \citep{Baines11}, hence providing an additional test. These {\it reference} fundamental stellar parameters are listed in Table \ref{tab:starparam}.

\subsection{Testing the Impact of the Optimization Procedure (TL Pipeline)} \label{sec:modeling_opt}

\subsubsection{Stellar Models, Input Physics, and Grid Computation} \label{sec:method_TL}

We use the stellar evolution code {\sc mesa} \citep[Modules for Experiments in Stellar Astrophysics, release version 12115;][]{2011ApJS..192....3P,2013ApJS..208....4P, 2015ApJS..220...15P,2018ApJS..234...34P,Paxton19} and the stellar oscillation code {\sc gyre} \citep[v5.1;][]{2013MNRAS.435.3406T,Townsend18} to compute a grid of stellar models\footnote{The corresponding {\sc mesa} inlists are available on Zenodo under an open-source Creative Commons Attribution license: 
\dataset[doi:10.5281/zenodo.7737358]{\doi{10.5281/zenodo.7737358}}.}. We adopt the solar chemical mixture, $(Z/X)_{\odot}$ = 0.0181, provided by \citet{AGS09}. The {\sc mesa} $\rho$--$T$ tables, based on the 2005 update of the {\sc opal} equation of state tables \citep{2002ApJ...576.1064R}, are adopted and we use {\sc opal} opacities supplemented by the low-temperature opacities from \citet{2005ApJ...623..585F}. The {\sc mesa} `Eddington' photosphere is used for the set of boundary conditions for modeling the atmosphere, i.e., the opacity of the model atmosphere is specified by the temperature of the outermost mesh point of the interior model via a gray Eddington $T$--$\tau$ relation. The mixing-length theory of convection is applied, parameterized by $\alpha_{\rm MLT}$. We consider convective overshooting in the core, hydrogen-burning shell, and envelope. The exponential scheme by \citet{2000A&A...360..952H} is applied. The overshoot parameter is mass-dependent and follows the relation $f_{\rm{ov}} = [0.13\,M({\rm M}_\odot) - 0.098]/9.0$ \citep{2010ApJ...718.1378M}. For models with masses above $2.0\,{\rm M}_\odot$, we adopt a fixed $f_{\rm{ov}}$ of 0.018. For a smooth convective boundary, we also apply the {\sc mesa} predictive mixing scheme. The mass-loss rate on the RGB is characterized by a Reimers' efficiency parameter \citep{Reimers75} of $\eta = 0.2$, constrained by seismic targets in the old open clusters NGC~6791 and NGC~6819 \citep{Miglio12}. Atomic diffusion is only considered for models with masses below $1.1\,{\rm M}_\odot$ during the main-sequence phase (it is turned off when the central hydrogen fraction falls below 0.01).

We compute a grid of models with masses ranging from 0.76 to $2.20\,{\rm M}_\odot$ and a step size of $0.02\,{\rm M}_\odot$. Besides the stellar mass, $M$, there are three other independent model inputs, namely, the initial helium fraction, $Y_{\rm i}$, the initial metallicity, [Fe/H]$_{\rm i}$, and the mixing-length parameter, $\alpha_{\rm MLT}$. Model input ranges and step sizes are provided in Table \ref{tab:grid}. We evolve stellar evolutionary tracks from the Hayashi line and terminate them either when $\log g \leq 1.5\;{\rm dex}$ on the RGB or helium-core burning starts (corresponding to an increase in the core heavy-element fraction).

\begin{table*}[!t]
\begin{center}
\caption{Model grids: Input ranges, step sizes, and main differences in terms of the input physics.}
\label{tab:grid}
\begin{tabular}{l|ccc|ccc}
\toprule
Input Parameters & \multicolumn{3}{c}{TL Pipeline} & \multicolumn{3}{c}{JO Pipeline\tablenotemark{a}}\\
& From & To & Step & From & To & Step\\
\hline
$M$ (${\rm M}_{\odot}$)  & \phantom{$-$}0.76 & 2.20 &  0.02 & \phantom{$-$}1.2\phantom{0} & 2.0\phantom{0} & ---\\
${\rm [Fe/H]}_{\rm i}$ (dex) & $-0.5$\phantom{0} & 0.5\phantom{0} & 0.1\phantom{0} & $-0.4$\phantom{0} & 0.4\phantom{0} & ---\\
$Y_{\rm i}$ & \phantom{$-$}0.24 & 0.32 & 0.02 & \phantom{$-$}0.25 & 0.32 & ---\\
$\alpha_{\rm{MLT}}$ & \phantom{$-$}1.7\phantom{0} & 2.5\phantom{0} & 0.2\phantom{0} & \phantom{$-$}1.55 & 1.95 & ---\\
\hline
Chemical Mixture & \multicolumn{3}{c}{\cite{AGS09}} & \multicolumn{3}{c}{\cite{gs98}}\\
Overshooting & \multicolumn{3}{c}{Mass-dependent} & \multicolumn{3}{c}{None}\\
Model Atmosphere & \multicolumn{3}{c}{`Eddington'} & \multicolumn{3}{c}{Varies}\\
Surface Correction\tablenotemark{b} & \multicolumn{3}{c}{\cite{Ball14}} & \multicolumn{3}{c}{\cite{Roxburgh16}}\\
\hline
\end{tabular}
\end{center}
\tablenotetext{a}{\scriptsize Step sizes are undefined as stellar models were computed over a quasirandomly sampled mesh of input parameters.}
\tablenotetext{b}{\scriptsize Concerns the optimization procedure rather than the input physics.}
\end{table*}

\subsubsection{Optimization} \label{sec:opt_TL}

The fitting scheme is based on a maximum likelihood estimation approach and is described in detail in \citet{2020MNRAS.495.3431L}. We adopt the spectroscopic $T_{\rm eff}$ and $[{\rm Fe}/{\rm H}]$ as classical constraints. A luminosity constraint is also adopted, although only in the cases of KOI-3886 and $\iota$ Dra. KIC~8410637 is in an eclipsing binary, which can potentially give rise to a biased estimate of the absolute luminosity, that being the reason why we opt for not imposing a luminosity constraint for this star. The fitting scheme employs the two-term surface correction method of \citet{Ball14} and further considers a model systematic uncertainty, which is estimated as the median frequency difference between observations and the best-fitting model. Moreover, mode frequencies are re-weighted as a function of their frequency difference with respect to \numax\ when calculating the likelihood.

We base the detailed modeling on three alternative, nested sets of seismic constraints. The first of these sets only considers $\ell = 0$ modes; the second includes $\ell = 0$ and 2 modes, as well as the most p-like $\ell = 1$ modes; and the third set makes use of all observed mode frequencies (which include g-dominated modes). We refer to the three implementations above as methods `0' (radial modes), `P' (p-like modes), and `A' (all modes), respectively. The most p-like dipole mode per radial order is manually selected from Tables \ref{tab:freq_KOI-3886} and \ref{tab:freq_iotaDra}. Since these modes are p-dominated, they have relatively high amplitudes and large widths in the power spectrum. Moreover, their frequency pattern in an \'echelle diagram should exhibit a curvature similar to that of radial modes. We follow both these criteria in selecting the most p-like dipole modes.

\subsubsection{Results and Discussion} \label{sec:res_TL}

\begin{figure*}[!t]
	\centering
	\includegraphics[width=\textwidth,trim=1.2cm 1.0cm 0.0cm 0.0cm,clip]{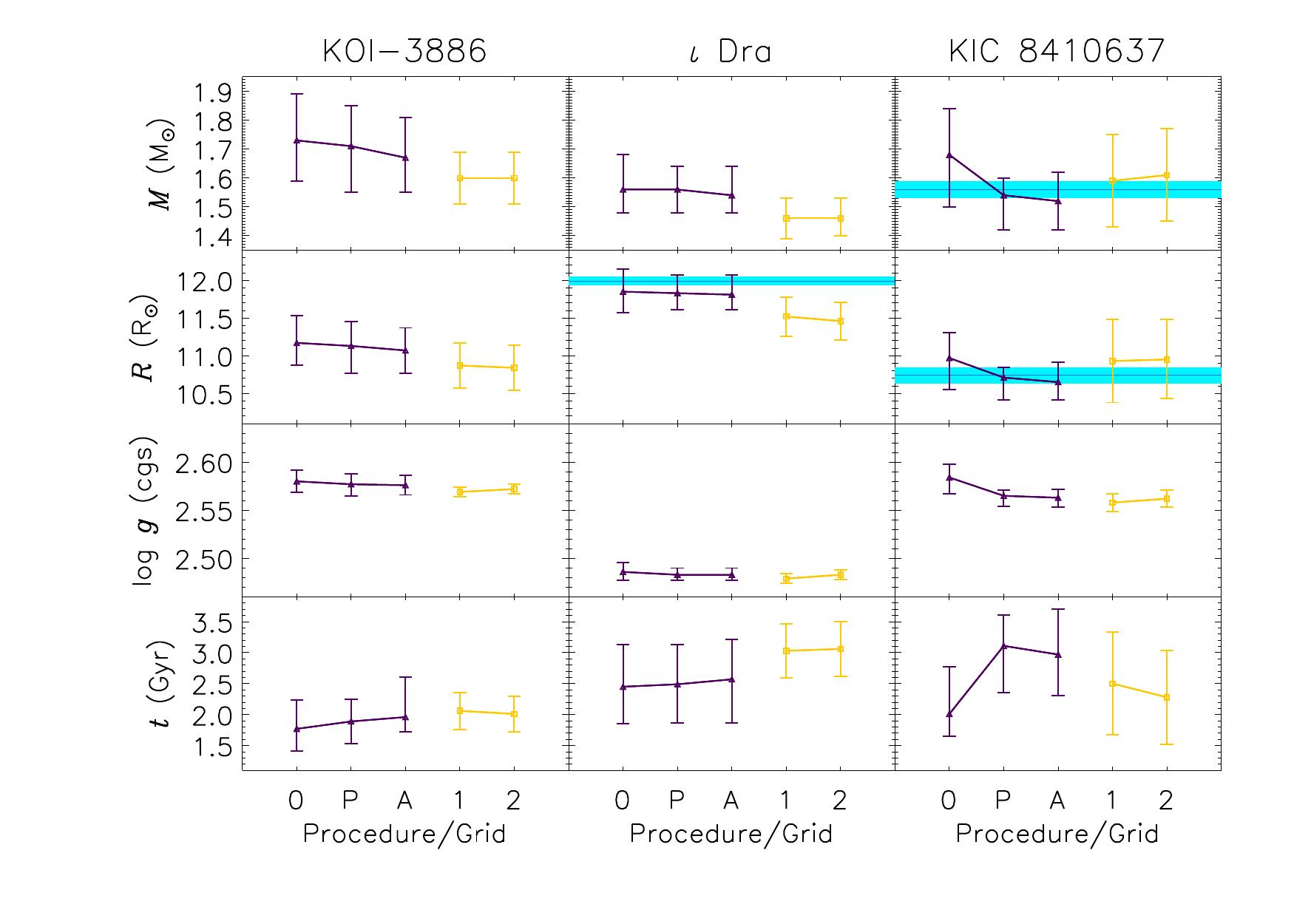}
	\caption{Estimated stellar parameters for KOI-3886, $\iota$ Dra, and KIC~8410637. Detailed modeling results from the TL (dark purple) and JO (yellow) Pipelines are shown. Procedures/Grids are labeled as `0', `P', `A' (cf.~Sect.~\ref{sec:modeling_opt}), and `1', `2' (cf.~Sect.~\ref{sec:modeling_phys}). Blue shaded areas represent the $1\sigma$ confidence intervals of the dynamical mass and radius of KIC~8410637, as well as of the interferometric radius of $\iota$ Dra.}
	\label{fig:properties}
\end{figure*}

\begin{figure*}[!t]
\centering
  \subfigure{%
  \includegraphics[width=.41\textwidth]{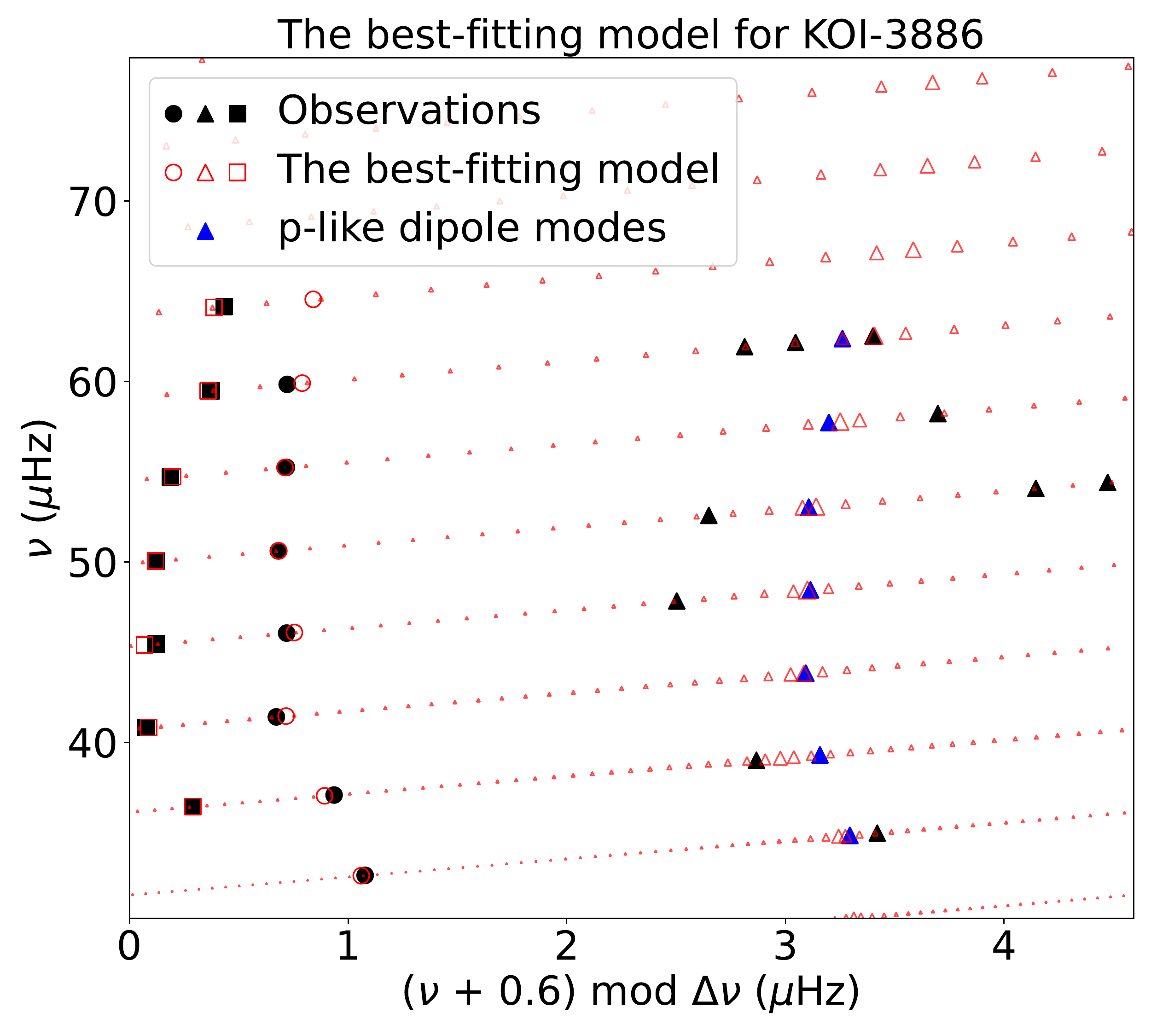}}
   \subfigure{%
  \includegraphics[width=.41\textwidth]{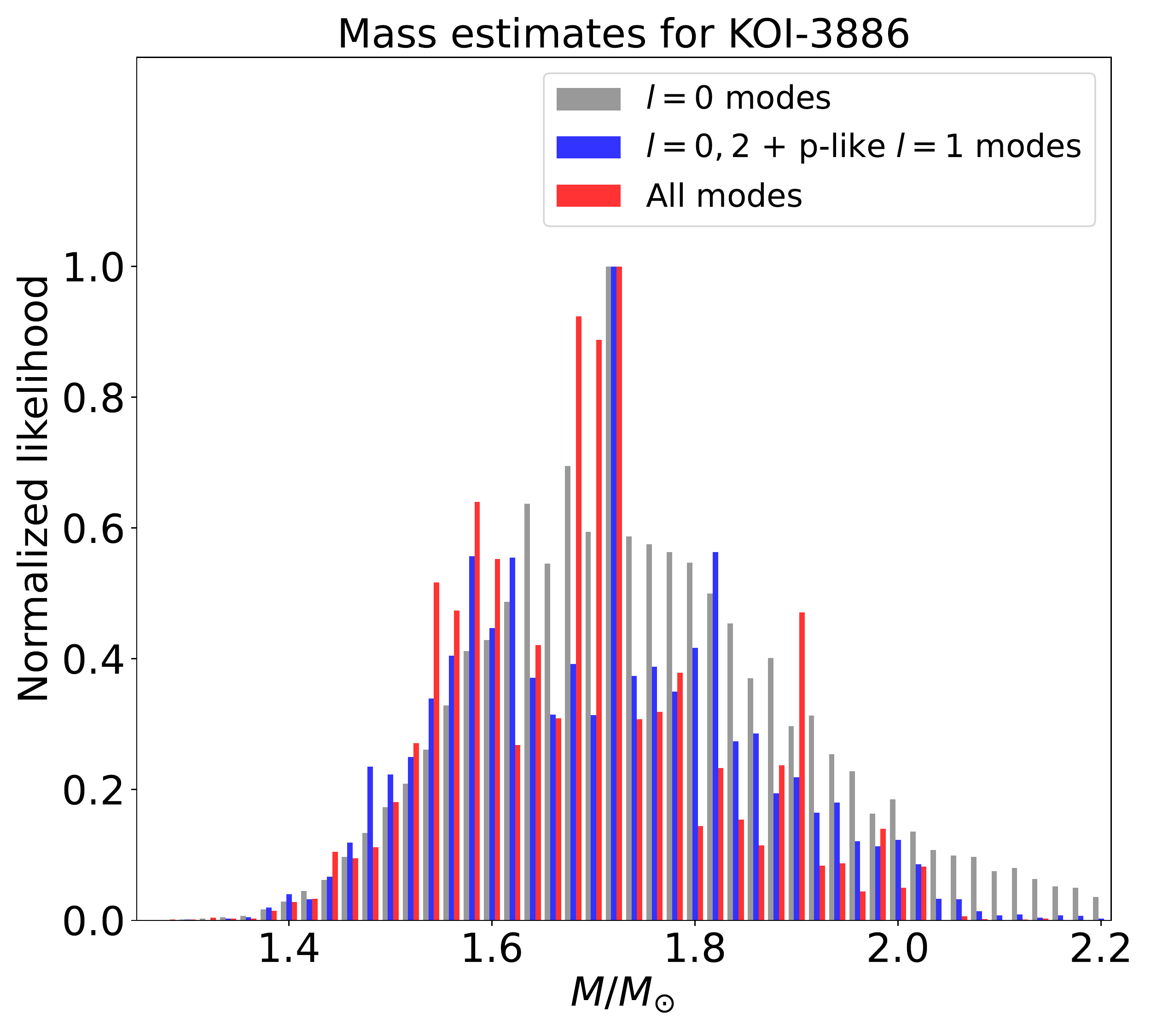}} 
  \subfigure{%
  \includegraphics[width=.41\textwidth]{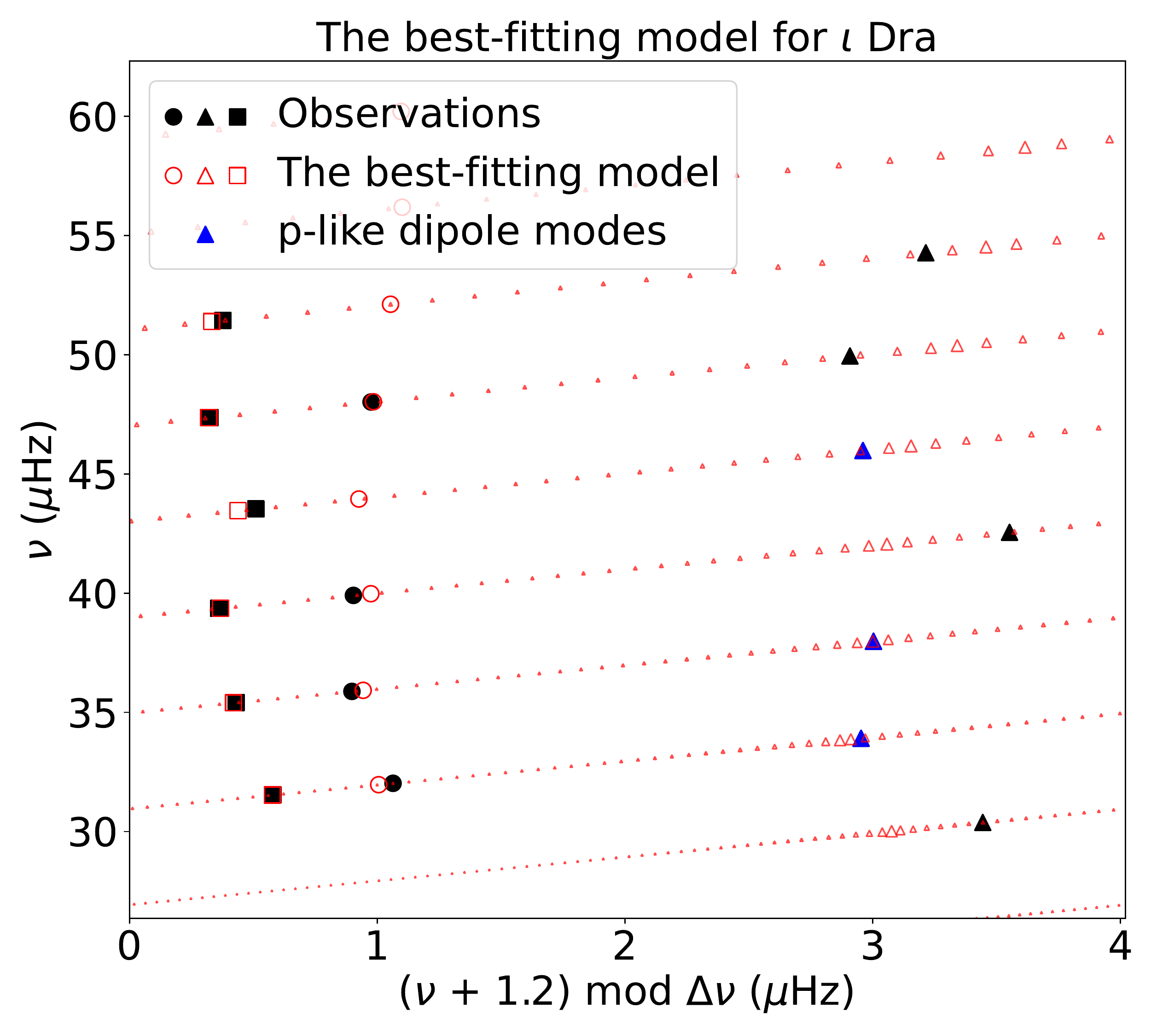}}
   \subfigure{%
  \includegraphics[width=.41\textwidth]{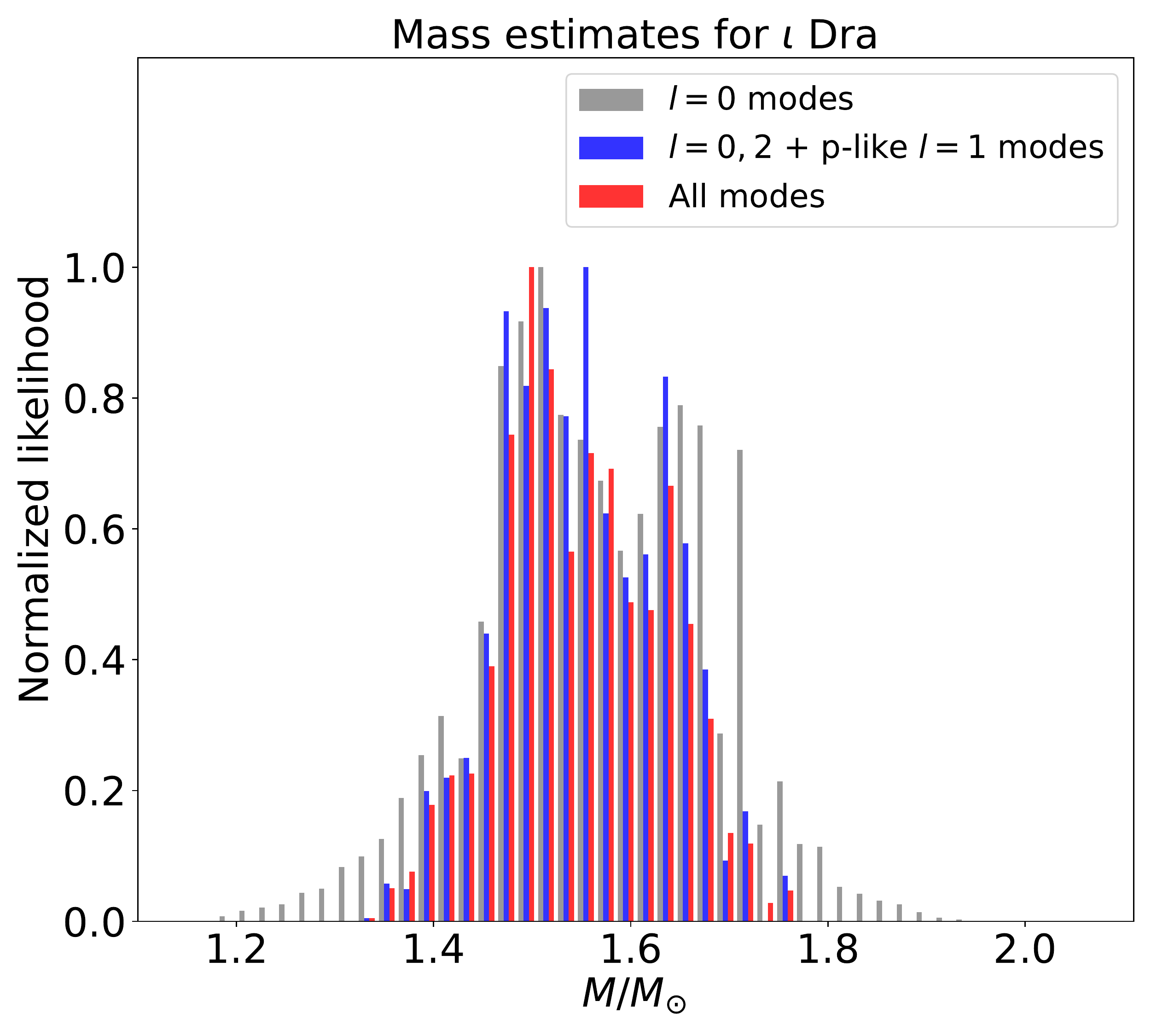}}   
    \subfigure{%
  \includegraphics[width=.41\textwidth]{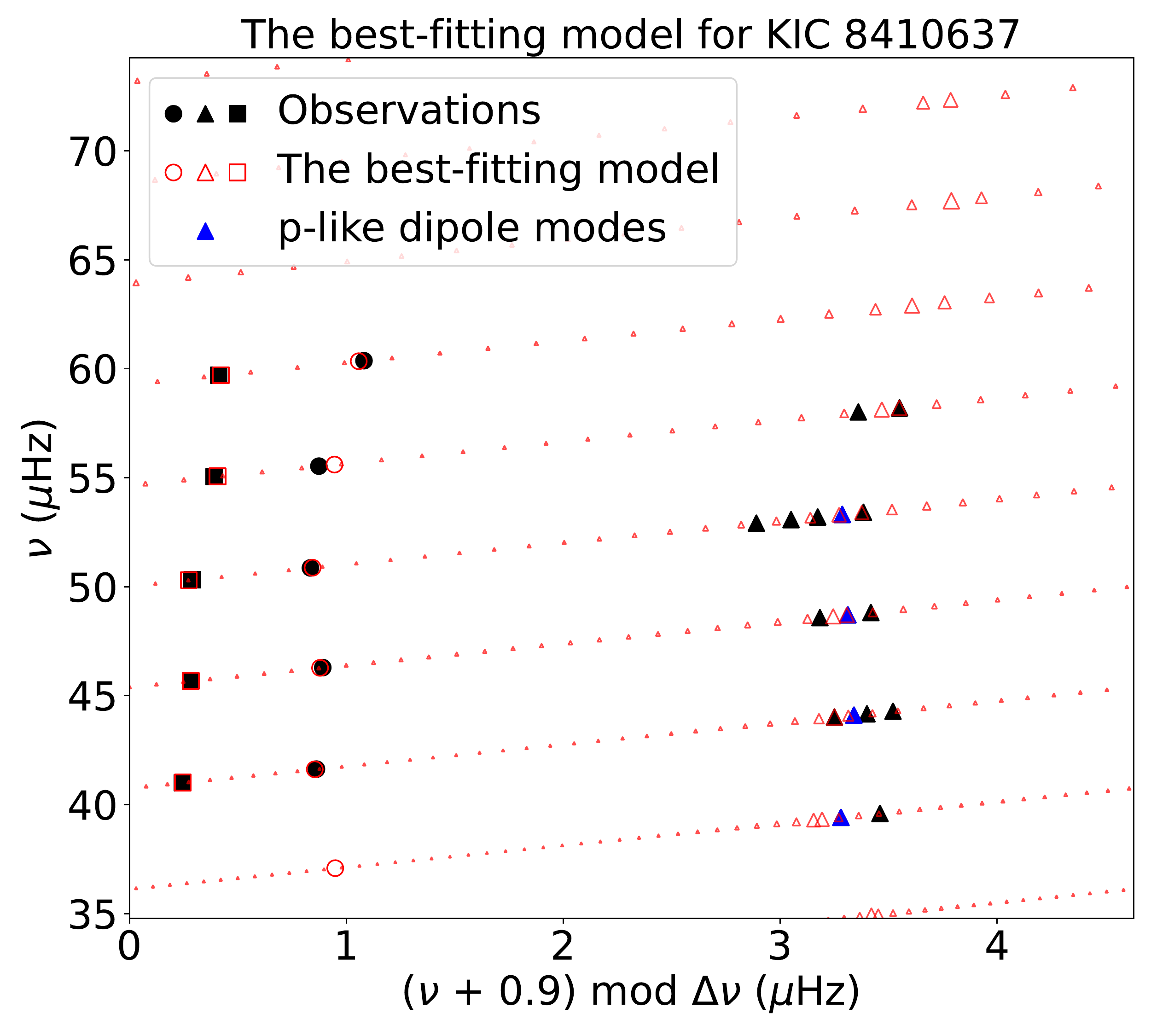}}
   \subfigure{%
  \includegraphics[width=.41\textwidth]{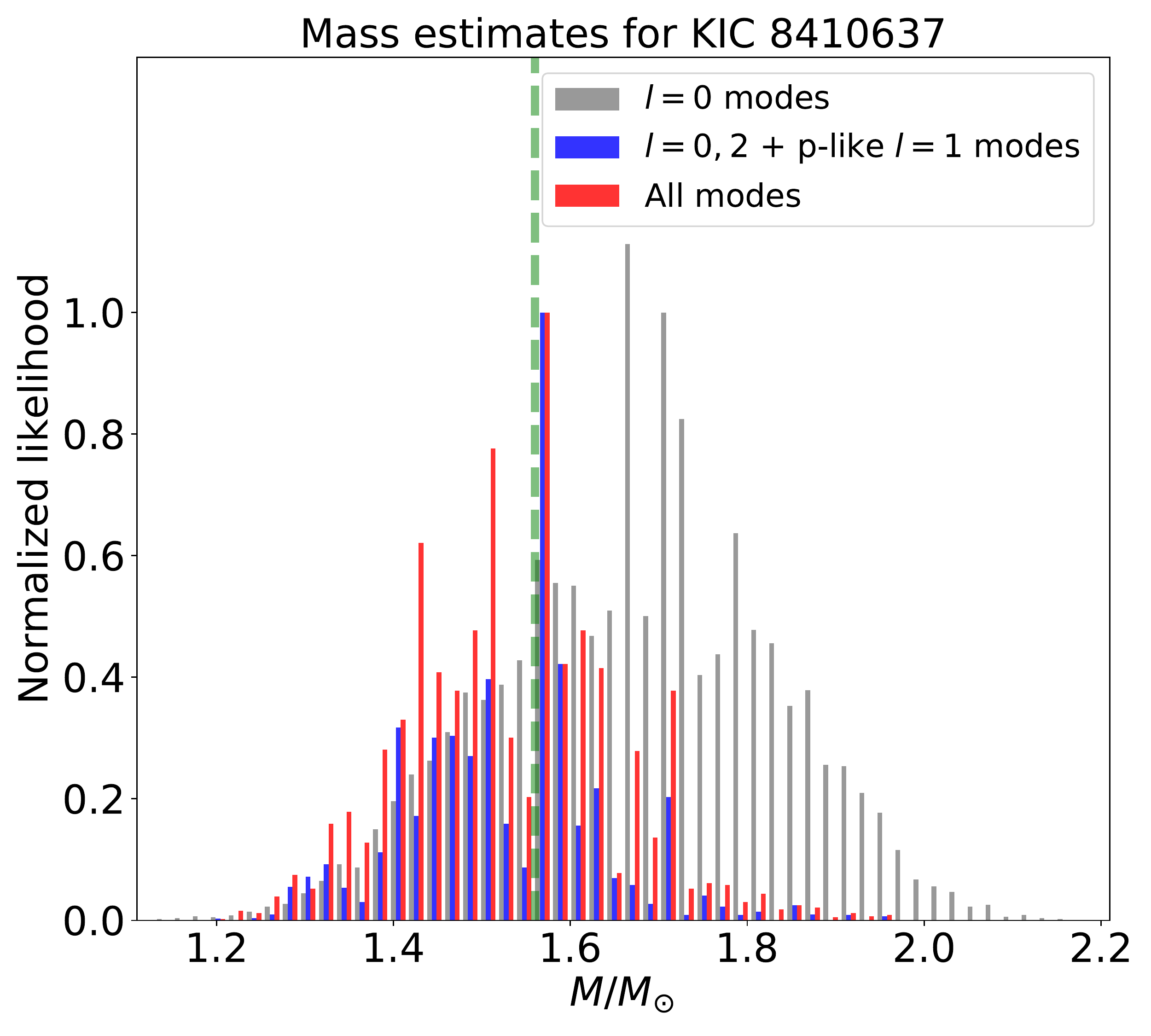}} 
  \caption{\emph{Left Column:} \'Echelle diagrams of the best-fitting models for (from top to bottom) KOI-3886, $\iota$ Dra, and KIC~8410637 constrained by all the observed mode frequencies (method `A'). Circles, triangles, and squares represent $\ell = 0$, 1, and 2 modes, respectively (with model frequencies shown as open symbols and observed frequencies as filled symbols). The most p-like observed dipole modes are rendered in dark blue. A range of mixed $\ell=1$ model frequencies are plotted with their symbol size scaled by the reciprocal of the mode inertia (larger size means that the mode is more p-dominated). The lower sparseness of the reported dipole mixed modes for KIC~8410637 compared to KOI-3886, the other \kepler\ target, has to do with the fact that peak-bagging of the former star used model frequencies to guide the identification of individual mixed modes \citep[cf.][]{Li18}. \emph{Right Column:} Probability distributions for the stellar mass estimated using each of the three optimization methods (`0' in gray, `P' in blue, and `A' in red). The vertical dashed line in the bottom panel represents the dynamical mass of KIC~8410637.}\label{fig:Tanda_ed}
\end{figure*}

Table \ref{tab:TL_est} lists the estimated stellar parameters (mass, radius, surface gravity, and age) stemming from each of the three optimization methods. This is complemented by Fig.~\ref{fig:properties}, where the dynamical mass and radius of KIC~8410637, as well as the interferometric radius of $\iota$ Dra are also represented. Figure \ref{fig:Tanda_ed} shows the best-fitting models (method `A') in an \'echelle diagram as well as the probability distributions (all three methods) for the stellar mass.

We are able to accurately retrieve (i.e., within the quoted measurement uncertainties) the available {\it reference} stellar parameters using each of the optimization methods (see Fig.~\ref{fig:properties}). The agreement with the dynamical solution for KIC~8410637 is particularly encouraging, especially bearing in mind the systematic overestimation of mass ($\sim15\%$) and radius ($\sim5\%$) for red giants by asteroseismic scaling relations reported by \citet{Gaulme16} \citep[see also][]{Brogaard18,Themessl18}. Inspection of the mass probability distributions (see Fig.~\ref{fig:Tanda_ed}) further reveals that the estimates returned by the different optimization methods are consistent within $1\sigma$ for all three stars (a statement that holds true if applied to the remaining stellar parameters, as can be seen in Fig.~\ref{fig:properties}).

Radial modes alone (method `0') are capable of constraining the fundamental parameters of the three stars in our sample with a precision of $2.4$--$3.5\%$, $6.4$--$10\%$, and $23$--$28\%$ on the radius, mass, and age, respectively. The inclusion of $\ell = 1$ and 2 modes in the fitting process (methods `P' and `A') leads only to a marginal gain in precision for both KOI-3886 and $\iota$ Dra. For KIC~8410637, the extra constraints provided by these modes lead to a more noticeable improvement not only in the precision (e.g., $5.8\%$ on the mass with method `P' versus $10\%$ with method `0'), but also in the accuracy of the stellar parameter estimates (e.g., a mass accurate within $1.3\%$ with method `P' versus $7.7\%$ with method `0'). Overall, a precision of $1.9$--$3.0\%$, $5.1$--$8.8\%$, and $19$--$25\%$ respectively on the radius, mass, and age is attained with method `P'. We note that these are relatively low-mass stars and thus have core conditions that do not vary significantly with stellar mass. This can be used to explain why the impact of including seismic indicators that probe the core (i.e., $\ell = 1$ g-dominated modes as well as the combination of $\ell = 0$ and 2 modes) is limited. This is illustrated, for instance, in figs.~4 and 5 of \citet{Lagarde16}, where the mass dependence of $\Delta\Pi_1$ along the RGB can be seen to almost vanish for $M\lesssim1.8\,{\rm M}_\odot$ \citep[see also fig.~4 of][]{Stello13}. Any residual mass dependence of $\Delta\Pi_1$ thus effectively becomes commensurate with the characteristic uncertainties, i.e., including statistical and systematic\footnote{For reference, the model systematic uncertainty for KIC~8410637, estimated as the median frequency difference between observations and the best-fitting model (see Sect.~\ref{sec:opt_TL}), takes the values $0.012$, $0.025$, and $0.009\:{\rm \mu Hz}$ for $\ell = 0$, 1, and 2 modes, respectively.} contributions, on the observed frequencies. This explanation holds true except perhaps where the mass approaches the degenerate transition. This is most noticeable in the case of KIC~8410637, for which a luminosity constraint was not imposed, and whose mass solutions above $1.8\,{\rm M}_\odot$ are removed upon adoption of core seismic constraints (see Fig.~\ref{fig:properties}).

We do not find significant differences between methods `P' and `A' in all three cases, i.e., access to dipole mixed modes seems not to improve our inference of the stellar parameters. As can be seen in the \'echelle diagrams in Fig.~\ref{fig:Tanda_ed}, g-dominated dipole mixed modes in evolved RGB stars are very densely spaced. If the gravity-mode period spacing is sufficiently small, essentially any arbitrary identification of the g-mode radial order of the sparse set of observationally available modes may be adopted to fit the dense forest of model mixed modes (except perhaps at the highest frequencies, for which the spacing becomes larger than the characteristic uncertainties on the observed frequencies). As a result, the posterior distributions for various stellar parameters become highly multimodal (see, e.g., the mass distribution for KOI-3886 corresponding to method `A' in Fig.~\ref{fig:Tanda_ed}), with each peak corresponding to a different choice of mode identification \cite[cf.~fig.~14 of][showing this phenomenon for less evolved subgiants]{Ong20}. This multimodality could in principle be alleviated with an a priori identification of the mixed modes, e.g., through the measurement of $\Delta\Pi_1$. Such measurements are not, however, available for these three stars.

Another aspect worth noting is that the precision achieved on the stellar parameters is similar across the three-star sample. Access to multi-year time-series photometry (resulting in higher-resolution power spectra) does not significantly improve on the detectability of the low-amplitude g-dominated mixed modes in evolved RGB stars. It is thus the shorter-lived p-like modes that mostly end up constraining the parameters of these stars. This explains why we obtain similar precision on the stellar parameters from detailed modeling for the multi-year \kepler\ targets KOI-3886 and KIC~8410637, and the multi-sector TESS target $\iota$ Dra, whose temporal coverage is substantially shorter (5 noncontiguous TESS sectors). We tested whether the fact that $\iota$ Dra is closer and brighter than the other two stars (see Table \ref{tab:starparam}), thus resulting in more precise non-seismic constraints, could be an important factor in this regard \citep[cf.][]{Stello22}. Having artificially degraded the precision on both $T_\mathrm{eff}$ and the luminosity so that they approximately match those for KOI-3886, methods `P' and `A' still return similarly precise stellar parameters compared to if the pristine uncertainties had been adopted.

\begin{table*}[!t]
\begin{center}
\caption{Estimated stellar parameters for KOI-3886, $\iota$ Dra, and KIC~8410637 (TL Pipeline).}
\label{tab:TL_est}
\renewcommand{\tabcolsep}{1mm}
\begin{tabular}{lcccccc}
\noalign{\smallskip}
\hline
\noalign{\smallskip}
Star & $M$ (${\rm M}_\odot$) & $R$ (${\rm R}_\odot$) &$\log g$ (cgs) & $t$ (Gyr) & Norm. RMS Dev.\tablenotemark{a} & Seismic Constraints\tablenotemark{b} \\
\hline
\noalign{\smallskip}
KOI-3886& $1.73^{+0.16}_{-0.14}$ & $11.17^{+0.36}_{-0.30}$ & $2.580^{+0.012}_{-0.011}$& $1.77^{+0.46}_{-0.36}$ & 0.20 (0.45) &0\\
\noalign{\smallskip}
        & $1.71^{+0.14}_{-0.16}$ & $11.13^{+0.32}_{-0.36}$ & $2.577^{+0.011}_{-0.012}$& $1.89^{+0.36}_{-0.36}$ & 0.04 (0.26) &P\\
\noalign{\smallskip}
        & $1.67^{+0.14}_{-0.12}$ & $11.07^{+0.30}_{-0.30}$ & $2.576^{+0.010}_{-0.010}$& $1.96^{+0.64}_{-0.24}$ & 0.20 (0.18) &A\\
\noalign{\smallskip}
\hline
\noalign{\smallskip}
$\iota$ Dra & $1.56^{+0.12}_{-0.08}$ & $11.85^{+0.30}_{-0.28}$ & $2.486^{+0.010}_{-0.009}$& $2.45^{+0.68}_{-0.60}$ & 0.12 (0.44) &0\\
\noalign{\smallskip}
        & $1.56^{+0.08}_{-0.08}$ & $11.83^{+0.24}_{-0.22}$ & $2.483^{+0.007}_{-0.006}$& $2.49^{+0.64}_{-0.62}$ & 0.09 (0.44) &P\\
\noalign{\smallskip}
        & $1.54^{+0.10}_{-0.06}$ & $11.81^{+0.26}_{-0.20}$ & $2.483^{+0.007}_{-0.006}$& $2.57^{+0.64}_{-0.70}$ & 0.13 (0.31) &A\\
\noalign{\smallskip}
\hline 
\noalign{\smallskip}
 KIC~8410637 & $1.68^{+0.16}_{-0.18}$ & $10.97^{+0.34}_{-0.42}$ & $2.584^{+0.014}_{-0.017}$& $2.01^{+0.76}_{-0.36}$ & 0.84 (0.82) &0\\
\noalign{\smallskip}
        & $1.54^{+0.06}_{-0.12}$ & $10.71^{+0.14}_{-0.30}$ & $2.565^{+0.006}_{-0.011}$& $3.11^{+0.50}_{-0.76}$ & 0.54 (0.59) &P\\
\noalign{\smallskip}
        & $1.52^{+0.10}_{-0.10}$ & $10.65^{+0.26}_{-0.24}$ & $2.563^{+0.009}_{-0.010}$& $2.97^{+0.74}_{-0.66}$ & 0.60 (0.61) &A\\
\noalign{\smallskip}
\hline 
\noalign{\smallskip}        
 & & & & & \phantom{$^\dagger$}0.31 (0.46)$^\dagger$ & \\
\noalign{\smallskip}
\hline
\end{tabular}
\end{center}
\tablenotetext{a}{Normalized RMS deviation about the mean ($d_{\rm norm}$; see Eq.~\ref{eq:dnorm}). Values outside (inside) brackets are computed considering mean parameter values across the set of procedures `0', `P', and `A' (all procedures/grids, i.e., `0', `P', `A', `1', and `2'). See Sect.~\ref{sec:interdisp} for details.}
\tablenotetext{b}{`0': $\ell=0$ modes only; `P': $\ell=0$ and 2 modes, as well as most p-like $\ell=1$ modes; `A': All observed modes.}
\tablenotetext{\dagger}{Average value, i.e., $\langle d_{\rm norm}\rangle$.}
\end{table*}

\subsection{Testing the Impact of the Input Physics: Model Atmosphere (JO Pipeline)} \label{sec:modeling_phys}

\subsubsection{Stellar Models, Input Physics, and Grid Computation} \label{sec:method_JO}

A different set of stellar models\footnote{The corresponding {\sc mesa} inlists are available on Zenodo under an open-source Creative Commons Attribution license: 
\dataset[doi:10.5281/zenodo.7737358]{\doi{10.5281/zenodo.7737358}}.} was generated with {\sc mesa} release version 12778 using the relative elemental abundances of \cite{gs98}, without element diffusion and convective overshooting, and allowing the initial helium abundances, initial metallicities, and mixing-length parameter to vary freely. A mass-loss rate characterized by a Reimers' efficiency parameter of $\eta = 0.2$ was adopted, as in the previous section. We retained stellar models at a constant temporal spacing of 0.2 Myr, starting from the point where $\nu^2\Delta\Pi_1 / \Delta\nu = 5$ (corresponding to roughly $\log g \sim 3.2$) until the tip of the RGB.

Two model grids (grids `1' and `2') were constructed in this manner, although with a different treatment of the model atmosphere. Even if both model grids use photospheric boundary conditions with respect to a gray atmosphere in the Eddington approximation, for grid `1' (or `spherical atmosphere') the model mesh was extended outwards to an optical depth of $\tau = 10^{-3}$ under spherical geometry (joined with a small plane-parallel section from $\tau = 10^{-4}$ to $\tau = 10^{-3}$), while in grid `2' (or `plane-parallel atmosphere') the photospheric boundary condition was integrated under plane-parallel geometry from an optical depth of $\tau = 10^{-3}$ to $\tau = {4 \over 3}$, where it was joined with the inner spherical mesh. The latter boundary condition is the same as that used in the previous section, except that here the atmospheric opacity is allowed to vary consistently with the local temperature and pressure in the atmosphere, rather than being held fixed to the outermost cell of the interior model.

For each choice of input physics, we then computed stellar models over a quasirandomly sampled mesh of initial parameters $\{M, [\mathrm{Fe/H}]_{\rm i}, Y_{\rm i}, \alpha_{\mathrm{MLT}}\}$ (see Table \ref{tab:grid}). These values were distributed uniformly over the intervals $M \in [1.2,2.0]\,{\rm M}_\odot$, $[\mathrm{Fe/H}]_{\rm i} \in [-0.4,0.4]\,\mathrm{dex}$, $Y_{\rm i} \in [0.25,0.32]$, and $\alpha_{\mathrm{MLT}} \in [1.55,1.95]$ by sampling over this parameter space with respect to a joint Sobol sequence, 4000 elements long. The initial mass and metallicity were twice as densely sampled as the other parameters, owing to the wide ranges spanned by these parameters.

\subsubsection{Optimization} \label{sec:opt_JO}

At high luminosities, the coupling between the p- and g-mode cavities in red giants is so weak that mode frequency measurement errors dominate over the characteristic frequency scales associated with mixed-mode coupling \citep[cf.~appendix A of][]{Ong21b}, as well as systematic errors caused by, e.g., an inappropriate surface-term correction. Furthermore, based on the findings of Sect.~\ref{sec:modeling_opt}, it is the most p-like modes that primarily contribute to the constraints on stellar parameters for evolved RGB stars. For these reasons, we hereafter restrict our attention to the most p-dominated modes, by means of which the notional pure p modes derived from stellar models are constrained. When the two mode cavities are treated as being decoupled, the numerical evaluation of these notional pure p modes has the benefit of becoming computationally far cheaper than for mixed modes.

Radial p-mode frequencies, and the frequencies of notional pure dipole and quadrupole p modes \citep[via the $\pi$-mode construction of][]{Ong20}, were computed for modes within $\pm 4\Delta\nu$ of $\nu_{\mathrm{max}}$. As in \cite{Ong21}, we applied the $\varepsilon_\ell$-matching algorithm of \cite{Roxburgh16} to yield a surface-independent discrepancy function, $\chi^2_\varepsilon$, on the internal structure from the mode frequencies. This algorithm operates by constructing diagnostic quantities out of both the model and observed mode frequencies, independently for each degree $\ell$. Agreement with the internal structure of the model, insensitive to the stellar surface, is achieved when these quantities collapse to a single function of frequency, in principle minimizing $\chi^2_\varepsilon$ when a nonparametric functional model is fitted against the data points. In order to allow the inclusion of the dipole modes in this procedure, the most p-like dipole mode per radial order was manually selected from Tables \ref{tab:freq_KOI-3886} and \ref{tab:freq_iotaDra}, and the associated frequency measurement error was inflated by adding in quadrature the local g-mode spacing, $\nu^2\Delta\Pi_1$, which was also estimated manually. While this mode selection was conducted independently of that in Sect.~\ref{sec:opt_TL}, almost exactly the same modes ended up being used here. Moreover, we chose to use the three lowest-frequency modes (without regard for degree) for the purpose of regularizing the $\varepsilon_\ell$-matching algorithm, as described in \cite{Ong21}.

For each model in the grid, we compute a likelihood function, $\mathcal{L}_i \sim \exp[-\chi_i^2 /  2]$, where the discrepancy function, $\chi_i^2$, is comprised of several terms:
\begin{itemize}
    \item[---] $\chi^2_\mathrm{glob}$, being the sum of error-normalized discrepancies for global parameters, namely, the classical spectroscopic quantities $T_\mathrm{eff}$ and $[\mathrm{Fe/H}]$, the luminosity (adopted only for KOI-3886 and $\iota$ Dra; cf.~Sect.~\ref{sec:opt_TL}), and $\nu_\mathrm{max}$, computed from models using the scaling relation.
    \item[---] $\chi^2_\varepsilon$, which is the reduced $\chi^2$ statistic returned by the nonparametric $\varepsilon_\ell$-matching algorithm of \cite{Roxburgh16}.
    \item[---] $\chi^2_\mathrm{reg}$, the regularization term describing the discrepancy for the three lowest-frequency modes. This term is downweighted by a factor of 4, so as not to unduly influence the shape of the posterior distribution.
\end{itemize}
Given a prior distribution over the parameter space of the grid, we are then able to define a posterior distribution, $p_i \propto \mathcal{L}_i w_i$, where $w_i$ is inversely proportional to the assumed prior distribution.

Finally, we estimate probability distributions for the stellar parameters using the Monte Carlo procedure described in \cite{Ong21}. In summary, the posterior mean for a given parameter, e.g., the stellar mass, is computed with respect to the likelihood function normalized by the sampling function of the grid (to impose the assumption of uniform priors) as $M \sim \sum_i p_i M_i$. This is done repeatedly with the likelihood function being re-evaluated under randomized perturbations to the observable constraints, as specified by their nominal measurement errors. The resulting distribution of the posterior means is then used to report the value and uncertainty of the parameter in question. While it would be prohibitively expensive --- computationally speaking --- to include perturbations to the mode frequencies, omitting their errors from this procedure has been shown, for main-sequence stars at least, not to appreciably affect the resulting posterior distributions, except that for the stellar age \citep[cf.][]{Cunha21}. In the case of red giants, however, their rapid evolution is such that their ages and masses are tightly correlated, and so this omission also leads to an underestimation of the uncertainties in parameters other than the age. For this reason, we instead report, for each parameter, the quadratic mean of two different error estimates: the $1\sigma$ quantiles of the distribution of the posterior means (the usual approach), as well as the posterior standard deviation associated with a single realization of the procedure (representing the frequency uncertainties).

\subsubsection{Results and Discussion} \label{sec:res_JO}

\begin{figure*}[htbp]
	\centering
	\includegraphics[width=\textwidth]{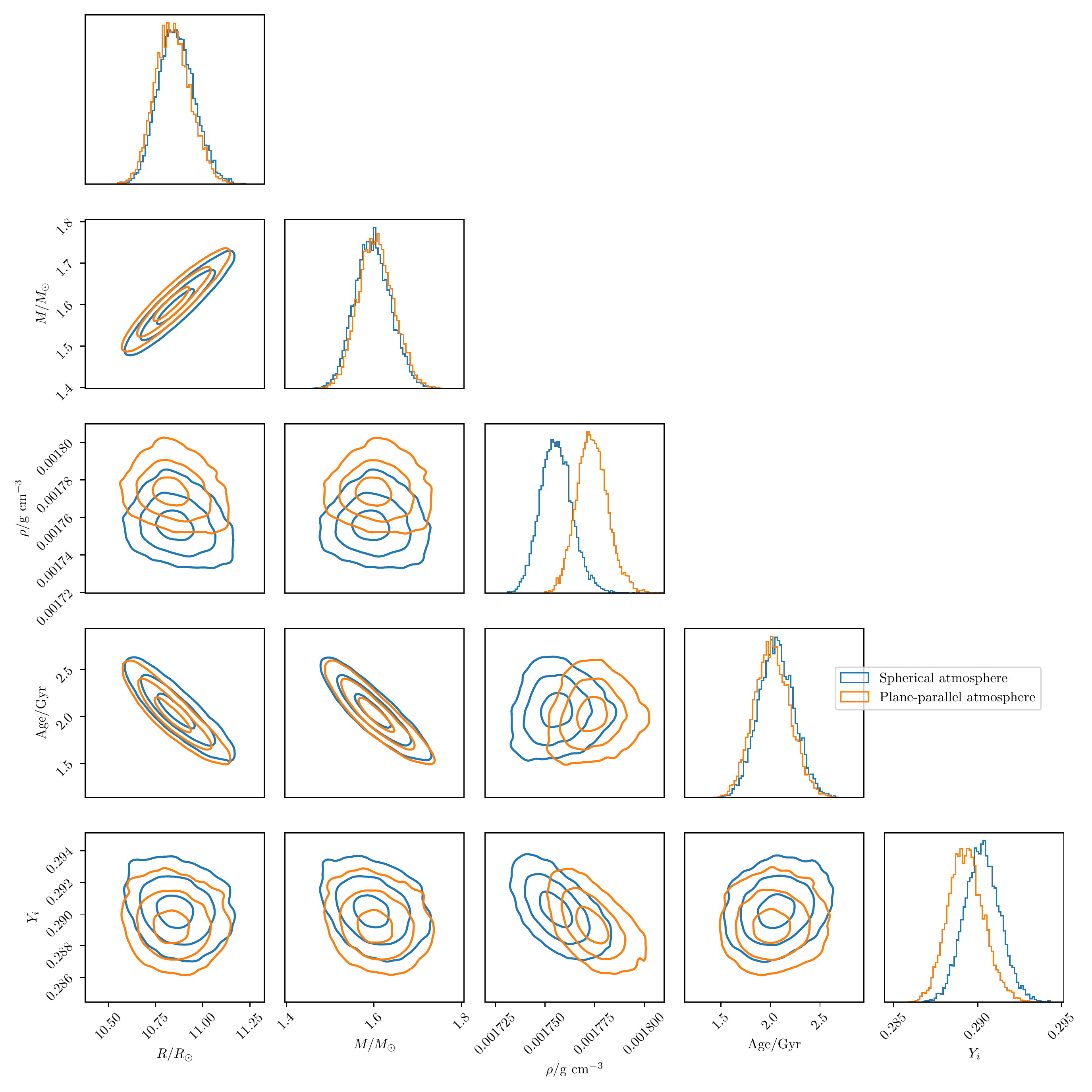}
	\caption{Joint posterior distribution of the stellar parameters for KOI-3886, showing a comparison between each choice of physics in the underlying grids of evolutionary models.}
	\label{fig:joint1}
\end{figure*}

\begin{figure*}[htbp]
	\centering
	\includegraphics[width=\textwidth]{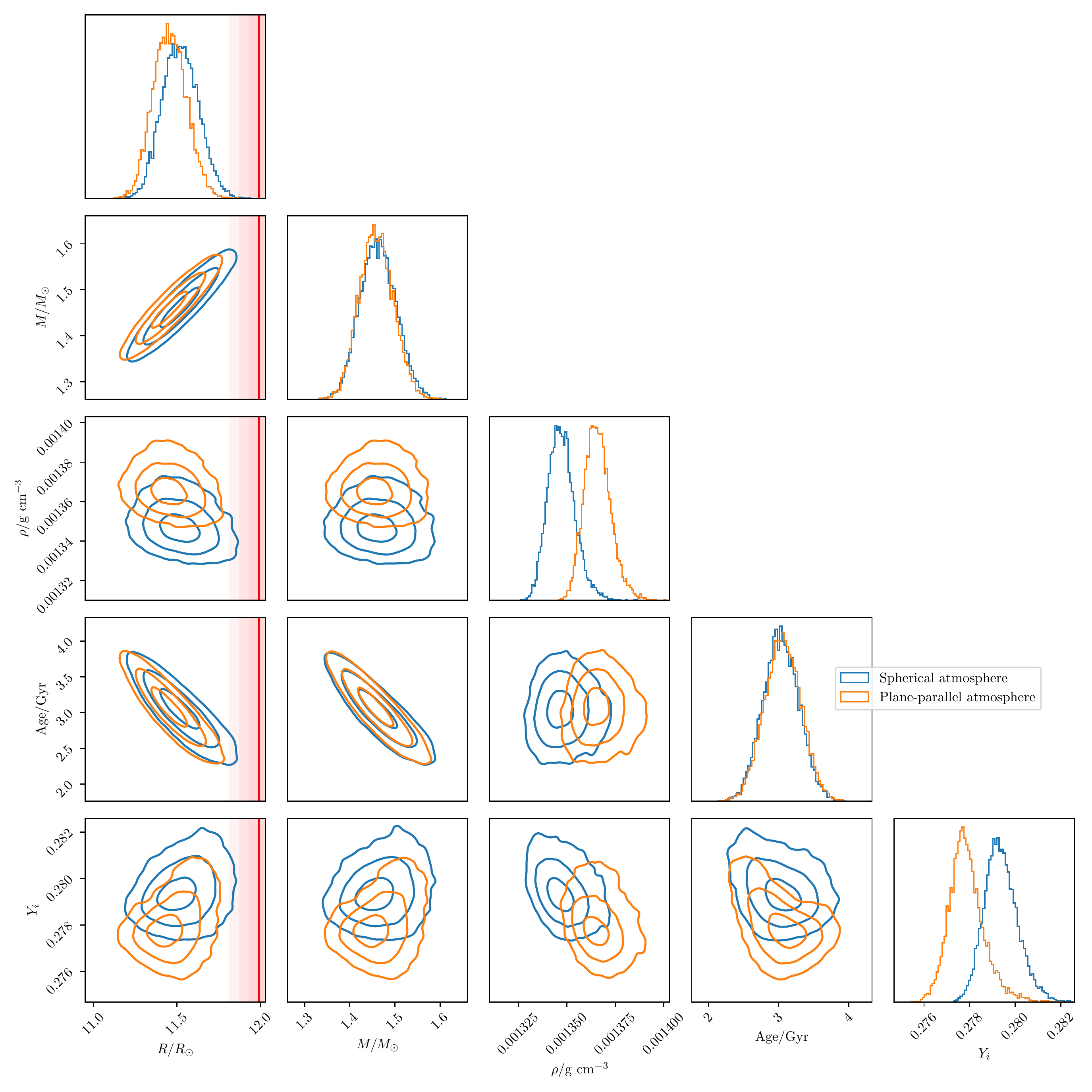}
	\caption{Joint posterior distribution of the stellar parameters for $\iota$ Dra, showing a comparison between each choice of physics in the underlying grids of evolutionary models. The interferometric radius from \cite{Baines11}, as well as its associated uncertainty ($1\sigma$, $2\sigma$, and $3\sigma$), are depicted by the vertical line and shaded region, respectively.}
	\label{fig:joint2}
\end{figure*}

\begin{figure*}[htbp]
	\centering
	\includegraphics[width=\textwidth]{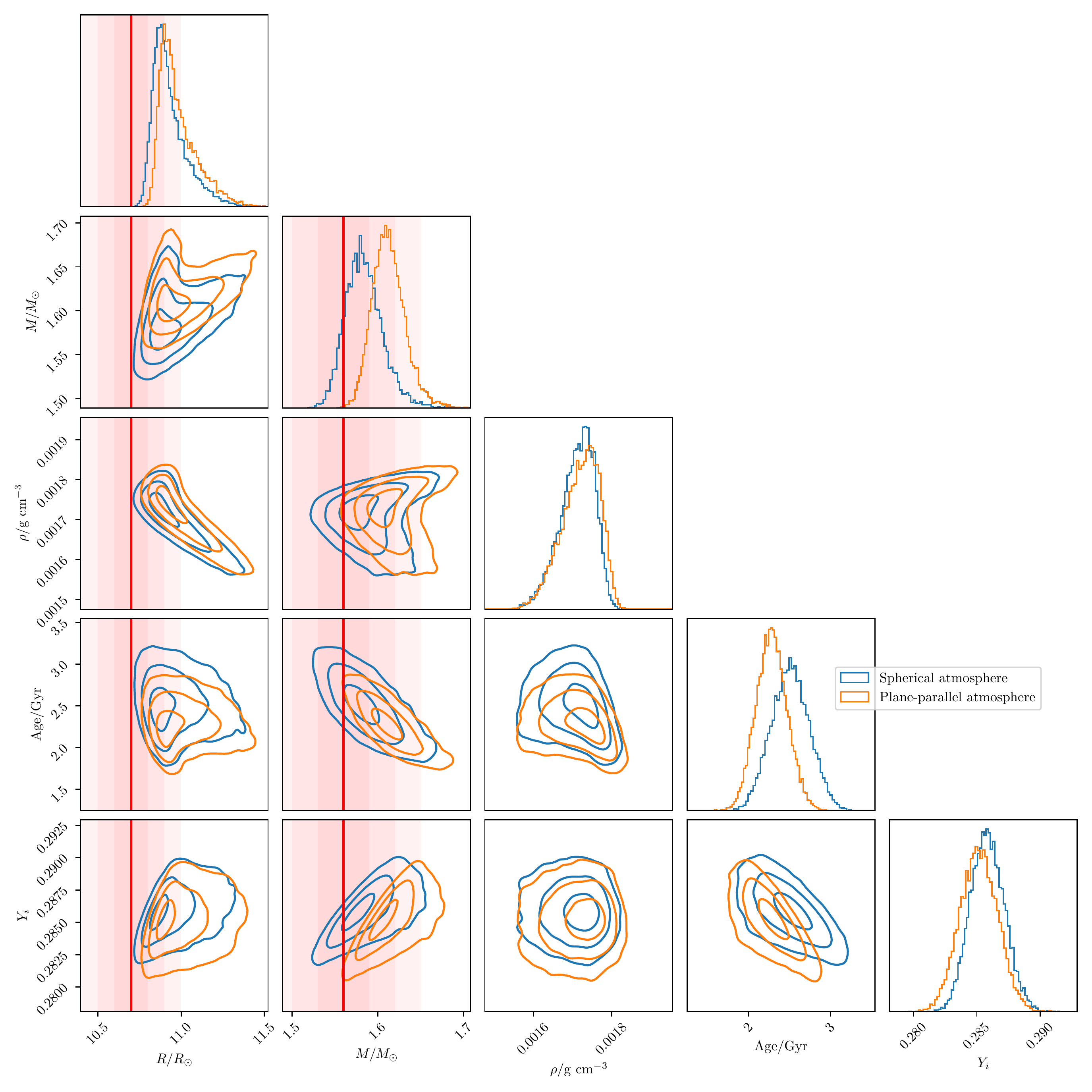}
	\caption{Joint posterior distribution of the stellar parameters for KIC~8410637, showing a comparison between each choice of physics in the underlying grids of evolutionary models. The dynamical mass and radius from \cite{2013A&A...556A.138F}, as well as their associated uncertainties ($1\sigma$, $2\sigma$, and $3\sigma$), are depicted by the vertical lines and shaded regions, respectively.}
	\label{fig:joint3}
\end{figure*}

We list in Table \ref{tab:JO_est} the mass, radius, surface gravity, and age estimates returned by the above procedure for each star, as applied to the two model grids (see also Fig.~\ref{fig:properties}). As in the previous section, we find very good agreement (i.e., within $1\sigma$) between the results of this exercise and both the dynamical mass and radius of KIC~8410637. The radius for $\iota$ Dra is in slight ($\sim2\sigma$) tension with the interferometric radius from \citet{Baines11}. We note that interferometric angular diameters can be subject to calibration biases \citep{White18}, in particular for measurements taken over a limited range of baselines or with partially resolved calibrators. Some of the systematic differences in angular diameters between stars observed with different instruments \citep{Tayar22} could account for the observed difference with respect to the asteroseismic radius. The Monte Carlo procedure also allows estimating both the marginal and joint posterior distributions of the stellar parameters, which we show in \cref{fig:joint1,fig:joint2,fig:joint3}. Note, following our discussion above, that the widths of the distributions returned by the bootstrapping procedure, which quantify the variations in the posterior mean under different realizations of the random error, are smaller than the reported uncertainties (these also include the posterior variances associated with individual realizations).

In \cref{fig:joint1,fig:joint2,fig:joint3}, results corresponding to both choices of the atmospheric boundary condition are represented by blue (grid `1' or `spherical atmosphere') and orange (grid `2' or `plane-parallel atmosphere') contours and histograms. We see marked differences in the inferred mean densities for KOI-3886 and $\iota$ Dra. These changes are commensurate with the small changes in $M$ and $R$. However, while the estimated masses and radii are not significantly changed, the changes in the mean densities are larger than the reported uncertainties. This is explained by the fact that the mean density is constrained (albeit indirectly, via the seismic data) with an order of magnitude more relative precision than $M$ and $R$ separately. On the other hand, this effect is not seen for KIC~8410637, for which a luminosity constraint was not adopted. Since $L$ is not imposed, changing the atmospheric boundary condition allows the best-fitting model to have potentially a different luminosity (and therefore mass and age) in order to better satisfy the very tight $\Delta\nu$ constraint.

Aside from these statistical considerations, there are also significant physical and methodological implications associated with changing the atmospheric boundary condition. In terms of the spectroscopic observables, it is well established that the choice of atmospheric boundary condition strongly determines the location of the RGB in the HR diagram associated with any given pair of $Y_{\rm i}$, the initial helium abundance, and $\amlt$, the mixing-length parameter. Accordingly, when the atmospheric boundary condition is changed, the values of $Y_{\rm i}$ and $\amlt$ that produce consistency with a fixed set of temperature and luminosity constraints must also be adjusted. Indeed, we see this happening in the bottom rows of \cref{fig:joint1,fig:joint2,fig:joint3}, which show the joint and marginal distributions of $Y_{\rm i}$, where the value of $Y_{\rm i}$ that best describes each star is modified between each choice of atmospheric boundary condition (this is particularly noticeable in the case of $\iota$ Dra).

Changing the atmospheric boundary condition also modifies the mode frequencies of a stellar model (inducing a numerical seismic surface term). Existing methodological comparisons of surface-term treatments for red giants \citep{Ball18,Jorgensen20,Ong21}, or even in general \citep{Basu18,Compton18,Nsamba18}, have usually considered the effects of different parameterizations of, or corrections for, modeling errors in stellar surfaces, under numerical experiments in which these modeling errors (arising from how the underlying set of stellar models are being generated) are kept the same. In this case, however, we have performed a converse experiment. We have maintained the use of a single algorithm to mitigate the asteroseismic surface term throughout --- the surface-independent $\varepsilon_\ell$-matching scheme of \cite{Roxburgh16} --- while changing the atmospheric boundary condition associated with the 1D evolutionary models. In particular, we have chosen a mitigation scheme that is designed to yield seismic constraints which are insensitive to the near-surface layers altogether, rather than attempting to correct their effects on the mode frequencies per se. Accordingly, the resulting systematics which we obtain originates from how the spectroscopic, rather than seismic, properties of the stellar models depend on the construction of their surface layers.

\begin{table*}[!t]
\begin{center}
\caption{Estimated stellar parameters for KOI-3886, $\iota$ Dra, and KIC~8410637 (JO Pipeline).}
\label{tab:JO_est}
\renewcommand{\tabcolsep}{1mm}
\begin{tabular}{lcccccc}
\noalign{\smallskip}
\hline
\noalign{\smallskip}
Star & $M$ (${\rm M}_\odot$) & $R$ (${\rm R}_\odot$) &$\log g$ (cgs) & $t$ (Gyr) & Norm. RMS Dev.\tablenotemark{a} & Model Grid\tablenotemark{b} \\
\hline
KOI-3886 & $1.60 ^ {+0.09} _ {-0.09}$ & $10.87 ^ {+0.30} _ {-0.30}$ & $2.569 ^ {+0.005} _ {-0.005}$ & $2.06 ^ {+0.30} _ {-0.30}$ & 0.16 (0.72) & 1\\ \noalign{\smallskip}
 & $1.60 ^ {+0.09} _ {-0.09}$ & $10.84 ^ {+0.30} _ {-0.30}$ & $2.572 ^ {+0.005} _ {-0.005}$ & $2.01 ^ {+0.29} _ {-0.29}$ & 0.16 (0.55) & 2\\ \noalign{\smallskip}
\hline
$\iota$ Dra & $1.46 ^ {+0.07} _ {-0.07}$ & $11.52 ^ {+0.26} _ {-0.26}$ & $2.479 ^ {+0.005} _ {-0.005}$ & $3.03 ^ {+0.44} _ {-0.44}$ & 0.21 (0.74) & 1\\ \noalign{\smallskip}
 & $1.46 ^ {+0.07} _ {-0.06}$ & $11.46 ^ {+0.25} _ {-0.25}$ & $2.483 ^ {+0.005} _ {-0.005}$ & $3.06 ^ {+0.44} _ {-0.44}$ & 0.21 (0.74) & 2\\ \noalign{\smallskip}
\hline
KIC~8410637 & $1.59 ^ {+0.16} _ {-0.16}$ & $10.93 ^ {+0.55} _ {-0.55}$ & $2.558 ^ {+0.009} _ {-0.009}$ & $2.50 ^ {+0.83} _ {-0.83}$ & 0.13 (0.48) & 1\\ \noalign{\smallskip}
 & $1.61 ^ {+0.16} _ {-0.16}$ & $10.95 ^ {+0.53} _ {-0.52}$ & $2.562 ^ {+0.009} _ {-0.009}$ & $2.28 ^ {+0.76} _ {-0.76}$ & 0.14 (0.34) & 2\\ \noalign{\smallskip}
\hline 
\noalign{\smallskip}
 & & & & & \phantom{$^\dagger$}0.17 (0.60)$^\dagger$ & \\
\noalign{\smallskip} 
\hline
\end{tabular}
\end{center}
\tablenotetext{a}{Normalized RMS deviation about the mean ($d_{\rm norm}$; see Eq.~\ref{eq:dnorm}). Values outside (inside) brackets are computed considering mean parameter values across grids `1' and `2' (all procedures/grids, i.e., `0', `P', `A', `1', and `2'). See Sect.~\ref{sec:interdisp} for details.}
\tablenotetext{b}{`1': Spherical atmosphere; `2': Plane-parallel atmosphere. See Sect.~\ref{sec:method_JO} for details.}
\tablenotetext{\dagger}{Average value, i.e., $\langle d_{\rm norm}\rangle$.}
\end{table*}

\subsection{Intra- vs.~Inter-pipeline Dispersion} \label{sec:interdisp}

Since the TL and JO Pipelines employ different underlying grids of stellar models and analysis methodologies, a meaningful investigation of the source(s) of inter-pipeline systematics is not feasible. The exercises carried out in Sects.~\ref{sec:modeling_opt} and \ref{sec:modeling_phys} should instead be regarded as independent, i.e., we are interested in assessing the intra- as opposed to the inter-pipeline dispersion. 

We find it nonetheless instructive to roughly characterize the relative importance of the methodological choices made when operating each of these pipelines. To that end, we introduce a simple statistic, namely, the normalized root-mean-square (RMS) deviation about the mean:
\begin{equation}\label{eq:dnorm}
    d_{\rm norm} = \sqrt{\frac{1}{4} \sum_i \left(\frac{\theta_i-\mu_{\theta_i}}{\sigma_i}\right)^2} \, ,
\end{equation}
where $\theta_i$ represents each of the four estimated stellar parameters ($M$, $R$, $\log g$, and $t$), $\sigma_i$ is the associated uncertainty, and $\mu_{\theta_i}$ is the parameter's mean value across a set of procedures/grids. We compute $d_{\rm norm}$ for each star and procedure/grid combination and list it in Tables \ref{tab:TL_est} and \ref{tab:JO_est}. Values outside brackets are computed considering mean parameter values across the set of procedures/grids within the same pipeline (i.e., TL or JO). Values inside brackets are computed considering mean parameter values across all procedures/grids (i.e., TL and JO combined). Average values, $\langle d_{\rm norm}\rangle$, are also provided, which can be interpreted as proxies for the intra- and inter-pipeline systematics, respectively. Figure \ref{fig:properties} provides a visual aid.

Two features are worth noting. First, the inter-pipeline systematics dominates over the intra-pipeline systematics. This is more noticeable in the case of the JO Pipeline, whose intra-pipeline systematics is a factor of $\sim 2$ smaller than for the TL Pipeline. Second, the uncertainties on the stellar parameters returned by both pipelines are robust, in the sense that their magnitudes are larger than the inter-pipeline (and hence intra-pipeline) systematics, i.e., $\langle d_{\rm norm}\rangle < 1$.

\section{Summary and Conclusions} \label{sec:conclusions}

We conducted detailed asteroseismic modeling of the evolved RGB host stars KOI-3886 and $\iota$ Dra, making use of two independent and well-established pipelines. A third star, KIC~8410637, a member of an eclipsing binary, was also modeled and used as benchmark. These are typical seismic evolved RGB stars, in the sense that they are all characterized by a paucity of observed g-dominated dipole mixed modes, being further of relatively low mass. Multi-year \kepler\ time-series photometry is available for both KOI-3886 and KIC~8410637, whereas multi-sector TESS time-series photometry is available for $\iota$ Dra.

Making use of the TL Pipeline (Sect.~\ref{sec:modeling_opt}), we tested the impact of the optimization procedure by adopting different (albeit nested) sets of observed oscillation modes as seismic constraints, namely, radial modes (method `0'), p-dominated modes (method `P'), and all modes (method `A'). The main outcomes of this exercise are as follows:
\begin{itemize}
    \item[---] Radial modes alone (method `0') are capable of constraining the fundamental parameters of the three stars in our sample with a precision as high as $2.4\%$, $6.4\%$, and $23\%$ respectively on the radius, mass, and age. We note that, while the use of (uncorrected) asteroseismic scaling relations leads to comparable precision on the radius and mass, it nevertheless results in a significant overestimation of $\langle \Delta R/R \rangle = 10.7\% \pm 3.5\%$ and $\langle \Delta M/M \rangle = 19.9\% \pm 8.1\%$ \citep[cf.][]{Li22}.
    \item[---] Inclusion of $\ell = 1$ and 2 modes (methods `P' and `A') improves precision only marginally, with $1.9$--$3.0\%$ (radius), $5.1$--$8.8\%$ (mass), and $19$--$25\%$ (age) being reached when $\ell = 0$ and 2 modes, as well as the most p-like $\ell = 1$ modes, are used as constraints (i.e., method `P'). The limited impact of including seismic probes into the core of these relatively low-mass stars ($M\lesssim1.8\,{\rm M}_\odot$) can be explained by the fact that their core conditions do not vary significantly with stellar mass, as attested by the vanishing mass dependence of $\Delta\Pi_1$ along the RGB \citep[cf.][]{Stello13,Lagarde16}. However, for masses above this threshold, and limited classical constraints, core seismic constraints can have an impact. This is seen in the case of KIC~8410637, for which a luminosity constraint was not imposed. Here, $\ell = 0$ modes alone produce statistically significant solutions for masses above $1.8\,{\rm M}_\odot$, which are ruled out when the $\ell = 1$ modes are included in the fitting process.
    \item[---] Given the very small spacing of adjacent dipole mixed modes characteristic of evolved RGB stars, a sparse set of observed mixed modes is not able to provide extra constraints on the inferred stellar parameters. This happens because an arbitrary identification of the g-mode radial order of the observed modes may be adopted to fit the dense forest of model mixed-mode frequencies, resulting in posterior distributions for the stellar mass that are highly multimodal \cite[cf.][]{Ong20}. In principle, this multimodality may be alleviated by supplying an a priori identification of the radial orders of the observed mixed modes. In practice, however, doing so would require both accurate estimation of $\Delta\Pi_\ell$, as well as constraints on allowable values for the g-mode phase function, $\varepsilon_{\rm g}$, near $\nu_\mathrm{max}$. Such constraints may be provided, for example, by JWKB (Jeffreys--Wentzel--Kramers--Brillouin) analysis of the g-mode cavity when the star is significantly more or significantly less evolved than the RGB luminosity bump \citep{Pincon19}, or by homology relations in stars with degenerate helium cores given $\Delta\Pi_\ell$ \citep{Deheuvels2022}. However, more general considerations may require further theoretical investigation.
    \item[---] The very low surface amplitudes of g-dominated mixed modes in evolved RGB stars hinder their detection, even if allowing enough time to fully resolve such modes. Therefore, it is the shorter-lived p-like modes that mostly end up constraining the fundamental parameters of these stars. This explains why, despite the much shorter temporal coverage (and thus lower resolution of the corresponding power spectra) of TESS targets compared to multi-year \kepler\ observations, detailed modeling of the former can lead to similarly precise stellar parameters.
\end{itemize}

Next, using the JO Pipeline (Sect.~\ref{sec:modeling_phys}), we tested the impact of the adopted near-surface physics, namely, the atmospheric boundary condition. The main outcomes of this exercise are:
\begin{itemize}
\item[---] Changing the atmospheric boundary condition is known to substantially modify the asteroseismic surface term --- the differences in the mode frequencies between those of a star and a stellar model with identical interior structure, owing to modeling errors in the near-surface layers. Because these mode frequencies are measured far more precisely than the spectroscopic constraints, this surface term is typically assumed to dominate the systematic error when estimating stellar parameters, if left uncorrected or corrected inappropriately. However, we have shown that even when using a seismic constraint designed to be independent of the near-surface structure of stellar models, inferences of stellar parameters are still significantly dependent on choices of atmospheric physics, as the spectroscopic properties of the models are also modified by changing the atmospheric boundary condition.
\item[---] We see here that seismic estimates of mass and radius appear methodologically insensitive to the description of the near-surface layers. This occurs at the expense of substantially changing both the near-surface structure of the best-fitting models (i.e., changing the calibration constant in the scaling relation between $\Delta\nu$ and the mean density, which relies on homology arguments) and the values of associated parameters like the initial helium abundance (in tandem with the interaction with the spectroscopic constraints). This has been illustrated in this work by the atmospheric boundary condition, but other unknown/incomplete contributions of the near-surface physics may have a similar impact. Some alternatives are the modeling of the superadiabatic layer, incorporation of atmospheric convective overshooting, or incorrect atmospheric opacities. This issue implies that attempts to measure $Y_{\rm i}$ from seismic modeling of red giants, e.g., for the purpose of Galactic chemical evolution (measuring ${\rm d}Y/{\rm d}Z$ relations) or Galactic archaeology studies, are systematically impacted by the modeling of the surface layers.
\end{itemize}

Finally, we provide consolidated fundamental parameters for the evolved RGB host stars KOI-3886 and $\iota$ Dra. Tables \ref{tab:TL_est} (TL Pipeline) and \ref{tab:JO_est} (JO Pipeline) list the returned mass, radius, surface gravity, and age estimates. Since both pipelines employ different underlying grids of stellar models and analysis methodologies, we are able to roughly characterize the relative importance of the methodological choices made when operating each of the pipelines (Sect.~\ref{sec:interdisp}). The inter-pipeline systematics is seen to dominate over the intra-pipeline systematics. Moreover, the uncertainties on the stellar parameters returned by both pipelines are found to be robust, as their magnitudes are larger than the (dominant) inter-pipeline systematics.

\facilities{\kepler, TESS, \gaia.}

\software{{\sc kepseismic} (\url{https://archive.stsci.edu/prepds/kepseismic/}), {\sc kadacs} \citep{KADACS1,KADACS2,KADACS3}, {\sc forecaster} \citep{chen2017}, {\sc diamonds} \citep{DIAMONDS}, {\sc famed} \citep{FAMED}, {\sc Clumpiness} \citep{Clumpiness}, {\sc mesa} \citep{2011ApJS..192....3P,2013ApJS..208....4P, 2015ApJS..220...15P,2018ApJS..234...34P,Paxton19}, {\sc gyre} \citep{2013MNRAS.435.3406T,Townsend18}.}

\appendix
\counterwithin{figure}{section}
\renewcommand\thefigure{\thesection\arabic{figure}}
\counterwithin{table}{section}
\renewcommand\thetable{\thesection\arabic{table}}

\section{Frequency Lists and Peak-bagging} \label{sec:freqlists}

\begin{deluxetable}{clccc}
\tablecolumns{5}
\tablewidth{0pc}
\tablecaption{Mode frequencies extracted by {\sc famed} for KOI-3886.\label{tab:freq_KOI-3886}}
\tablehead{
\colhead{$n_{\rm p}$} & \colhead{$\ell$} & \colhead{Frequency ($\rm{\mu Hz}$)} & \colhead{1$\sigma$ Uncertainty ($\rm{\mu Hz}$)} & \colhead{$p_{\rm det}$\tablenotemark{a}}}
\startdata
6&0&32.629&0.047&1.000 \\
6&1$^{\dagger,\ddagger}$&34.846&0.021&0.996 \\
6&1&34.971&0.031&1.000 \\
6&2&36.434&0.065&1.000 \\
7&0&37.080&0.036&1.000 \\
7&1&39.011&0.044&1.000 \\
7&1$^{\dagger,\ddagger}$&39.302&0.042&--- \\
7&2&40.813&0.057&1.000 \\
8&0&41.408&0.065&--- \\
8&1$^{\dagger,\ddagger}$&43.830&0.041&--- \\
8&2&45.452&0.057&--- \\
9&0&46.049&0.057&--- \\
9&1&47.833&0.038&0.999 \\
9&1$^{\dagger,\ddagger}$&48.444&0.045&--- \\
9&2&50.045&0.070&--- \\
10&0&50.604&0.063&--- \\
10&1&52.573&0.030&1.000 \\
10&1$^{\dagger,\ddagger}$&53.030&0.041&--- \\
10&1&54.068&0.021&1.000 \\
10&1&54.397&0.035&1.000 \\
10&2&54.703&0.046&--- \\
11&0&55.233&0.068&--- \\
11&1$^{\dagger,\ddagger}$&57.715&0.061&--- \\
11&1&58.213&0.033&1.000 \\
11&2&59.482&0.061&1.000 \\
12&0&59.830&0.025&1.000 \\
12&1&61.922&0.006&0.993 \\
12&1&62.155&0.025&0.998 \\
12&1$^{\dagger,\ddagger}$&62.370&0.011&1.000 \\
12&1&62.509&0.015&0.997 \\
12&2&64.136&0.100&1.000 \\
\enddata
\tablenotetext{a}{A peak is tested against the noise only if its height in the smoothed power spectrum is lower than 10 times the local background level, otherwise it is automatically considered as detected (denoted as `---'). A detection probability ($p_{\rm det}$) is computed for each low-$S/N$ peak based on a Bayesian model comparison, peaks being deemed significant by {\sc famed} if $p_{\rm det} \ge 0.993$. See sect.~5.3 of \citet{FAMED} for details.}
\tablenotetext{\dagger}{\small Adopted as p-like dipole modes in Sect.~\ref{sec:opt_TL}.}
\tablenotetext{\ddagger}{\small Adopted as p-like dipole modes in Sect.~\ref{sec:opt_JO}.}
\end{deluxetable}

\begin{deluxetable}{clcccc}
\tablecolumns{6}
\tablewidth{0pc}
\tablecaption{Mode frequencies extracted by {\sc famed} for $\iota$ Dra.\label{tab:freq_iotaDra}}
\tablehead{
\colhead{$n_{\rm p}$} & \colhead{$\ell$} & \colhead{Frequency ($\rm{\mu Hz}$)} & \colhead{1$\sigma$ Uncertainty ($\rm{\mu Hz}$)} & \colhead{$p_{\rm det}$\tablenotemark{a}} & \colhead{List\tablenotemark{b}}}
\startdata
6&1&30.384&0.028&0.997&Min. \\
6&2&31.538&0.133&1.000&Min. \\
7&0&32.024&0.024&0.998&Min. \\
7&1$^{\dagger,\ddagger}$&33.913&0.088&1.000&Min. \\
7&2&35.410&0.039&---&Min. \\
8&0&35.878&0.035&---&Min. \\
8&1$^{\dagger,\ddagger}$&37.983&0.035&---&Min. \\
8&2&39.361&0.072&---&Min. \\
9&0&39.904&0.049&---&Min. \\
9&1&42.078&0.024&---&Max. \\
9&1&42.552&0.016&---&Min. \\
9&1&43.063&0.011&---&Max. \\
9&2&43.530&0.107&---&Min. \\
10&0&43.925&0.016&---&Max. \\
10&1$^{\dagger,\ddagger}$&45.980&0.027&---&Min. \\
10&2&47.364&0.072&---&Min. \\
11&0&48.015&0.038&---&Min. \\
11&1$^\ddagger$&49.948&0.027&---&Min. \\
11&1&50.420&0.022&0.997&Max. \\
11&2&51.436&0.025&---&Min. \\
12&1&54.274&0.021&0.999&Min. \\
12&2&55.202&0.100&0.999&Max. \\
13&0&55.565&0.025&0.994&Max. \\
\enddata
\tablenotetext{a}{See footnote to Table \ref{tab:freq_KOI-3886}.}
\tablenotetext{b}{\small `Min.' = Belongs to the minimal list; `Max.' = Belongs to the maximal list (but not to the minimal list). Only modes in the minimal list are subject to detailed modeling. See Sect.~\ref{sec:peak_bagging} for a definition of both lists.}
\tablenotetext{\dagger}{\small Adopted as p-like dipole modes in Sect.~\ref{sec:opt_TL}.}
\tablenotetext{\ddagger}{\small Adopted as p-like dipole modes in Sect.~\ref{sec:opt_JO}.}
\end{deluxetable}

\begin{figure*}[!t]
    \centering
    \includegraphics[width=\linewidth,trim=0cm 0cm 0cm 0cm,clip]{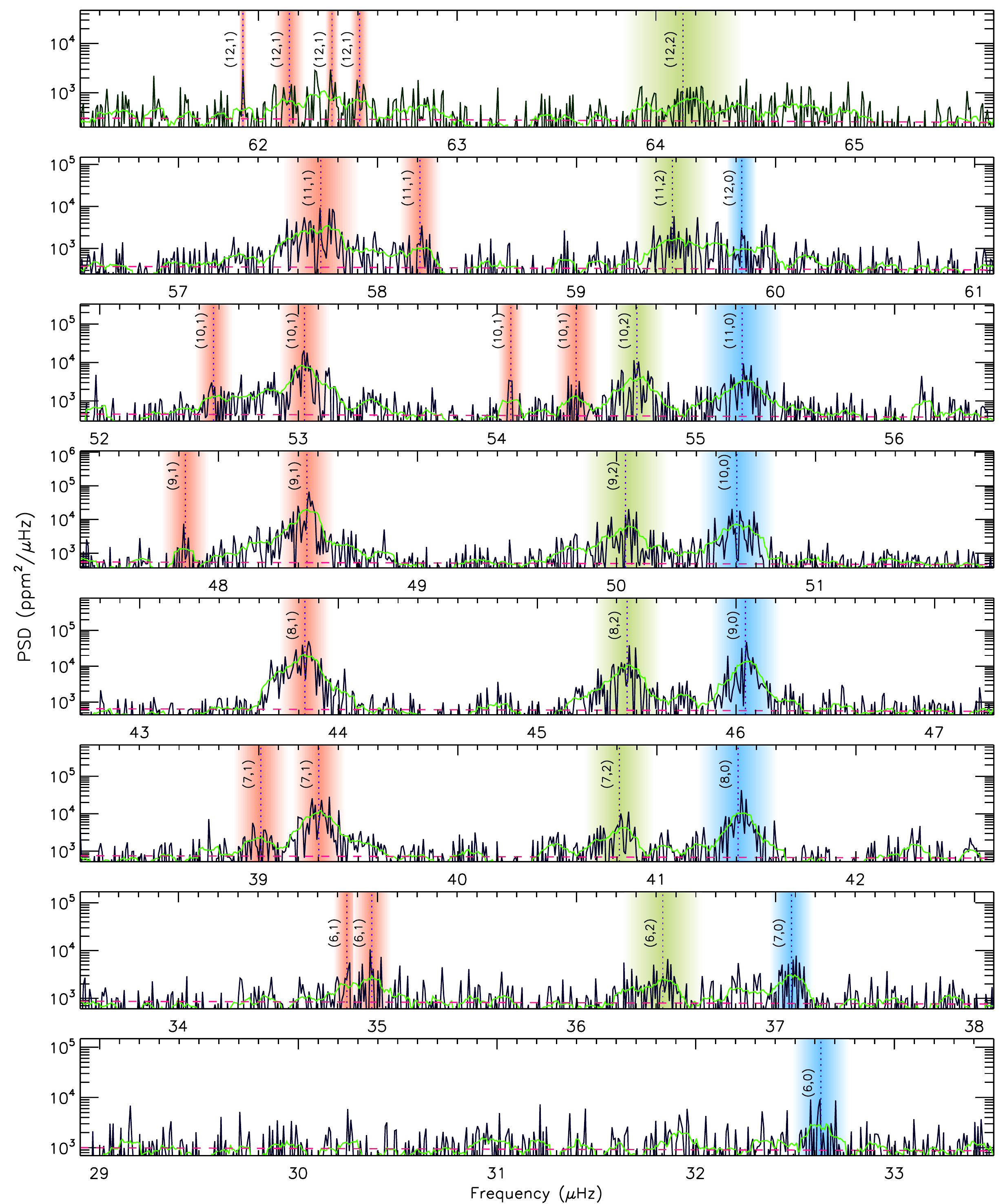}
    \caption{Stacked PSD of KOI-3886 showing the outcome of the peak-bagging process using {\sc famed}. The green curve is a smoothing of the power density by an amount equivalent to the average radial-mode linewidth. The sloping dashed line represents the local background. Extracted individual mode frequencies are tagged according to their pressure radial order ($n_{\rm p}$) and angular degree ($\ell$), with color bands indicating their $3\sigma$ uncertainties.}
    \label{fig:freq_KOI-3886}
\end{figure*}

\begin{figure*}[!t]
    \centering
    \includegraphics[width=\linewidth,trim=0.4cm 2.8cm 0.4cm 2.6cm,clip]{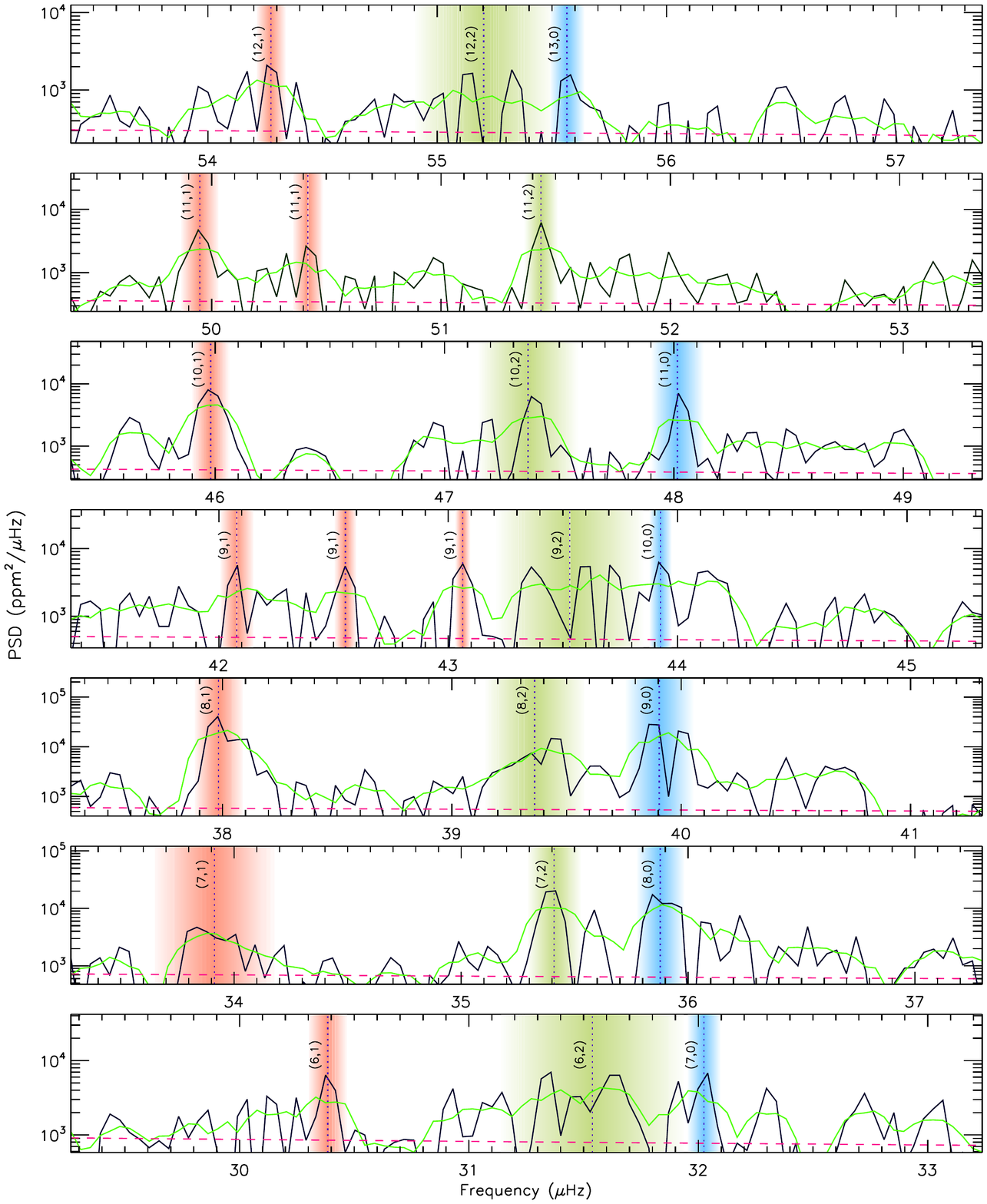}
    \caption{Stacked PSD of $\iota$ Dra showing the outcome of the peak-bagging process using {\sc famed}. Plot style similar to that of Fig.~\ref{fig:freq_KOI-3886}. The lower resolution of the TESS power spectrum of $\iota$ Dra is readily noticeable.}
    \label{fig:freq_iotaDra}
\end{figure*}

\clearpage
\bibliography{main}{}
\bibliographystyle{aasjournal}

\begin{acknowledgments}
This paper includes data collected by the \kepler\ and TESS missions. Funding for the \kepler\ and TESS missions is provided by the NASA Science Mission Directorate. This work was supported by Funda\c c\~ao para a Ci\^encia e a Tecnologia (FCT) through research grants UIDB/04434/2020 and UIDP/04434/2020. This work was supported by FCT through national funds (PTDC/FIS-AST/30389/2017) and by FEDER through COMPETE2020 (POCI-01-0145-FEDER-030389). T.L.C.~is supported by FCT in the form of a work contract (CEECIND/00476/2018). T.L.~is supported by the European Research Council (ERC) under the European Union's Horizon 2020 research and innovation programme (CartographY; Grant Agreement no.~804752). J.M.J.O.~acknowledges partial support by NASA grant 80NSSC19K0374, awarded to Sarbani Basu, and from NASA through the NASA Hubble Fellowship grant HST-HF2-51517.001 awarded by the Space Telescope Science Institute, which is operated by the Association of Universities for Research in Astronomy, Incorporated, under NASA contract NAS5-26555. We thank the Yale Center for Research Computing for guidance and use of the research computing infrastructure. M.S.C.~is supported by FCT in the form of a work contract (CEECIND/02619/2017). T.R.B.~was supported by the Australian Research Council (DP210103119). S.N.B.~and R.A.G.~acknowledge support from the PLATO CNES grant. M.D.~is supported by national funds through FCT in the form of a work contract. D.H.~acknowledges support from the Alfred P.~Sloan Foundation and the National Aeronautics and Space Administration (80NSSC19K0597, 80NSSC21K0652). C.J.~is supported by a grant from the Max Planck Society to prepare for the scientific exploitation of the PLATO mission. C.K.~acknowledges support from Erciyes University Scientific Research Projects Coordination Unit through grant DOSAP MAP-2020-9749. J.L.-B.~acknowledges financial support received from the ``la Caixa'' Foundation under Marie Sk\l{}odowska-Curie fellowship with code LCF/BQ/PI20/11760023 and the Spanish State Research Agency (AEI) projects PID2019-107061GB-C61 and MDM-2017-0737. S.M.~acknowledges support from the Spanish Ministry of Science and Innovation (MICINN) with the Ram\'on y Cajal fellowship no.~RYC-2015-17697 and grant no.~PID2019-107187GB-I00, and through AEI under the Severo Ochoa Centres of Excellence Programme 2020--2023 (CEX2019-000920-S). A.S.~is supported by grants PID2019-108709GB-I00 of the Spanish MICINN and Mar\'{i}a de Maeztu CEX2020-001058-M.
\end{acknowledgments}

\end{document}